\documentclass[prb,twocolumn,floatfix]{revtex4}

\usepackage{amsfonts}
\usepackage{amsmath}
\usepackage{verbatim}
\usepackage[dvips]{graphicx}
\usepackage[dvips]{color}
\usepackage{subfigure}
\usepackage{array}

\begin{document}

\title{Vortex states in a non-Abelian magnetic field}
\author{Predrag Nikoli\'c$^{1,2}$}
\affiliation{$^1$Department of Physics and Astronomy,\\George Mason University, Fairfax, VA 22030, USA}
\affiliation{$^2$Institute for Quantum Matter at Johns Hopkins University, Baltimore, MD 21218, USA}
\date{\today}

\begin{abstract}

A type-II superconductor survives in an external magnetic field by admitting an Abrikosov lattice of quantized vortices. This is an imprint of the Aharonov-Bohm effect created by the Abelian U(1) gauge field. The simplest non-Abelian analogue of such a gauge field, which belongs to the SU(2) symmetry group, can be found in topological insulators. Here we discover a superconducting ground state with a lattice of SU(2) vortices in a simple two-dimensional model that presents an SU(2) ``magnetic'' field (invariant under time-reversal) to attractively interacting fermions. The model directly captures the correlated topological insulator quantum well, and approximates one channel for instabilities on the Kondo topological insulator surface. Due to its simplicity, the model might become amenable to cold atom simulations in the foreseeable future. The vitality of low-energy vortex states born out of SU(2) ``magnetic'' fields is promising for the creation of incompressible vortex liquids with non-Abelian fractional excitations.

\end{abstract}

\maketitle

\section{Introduction}

The ubiquitous Rashba spin-orbit coupling, found in two-dimensional electron systems and topological insulators (TIs), is equivalent to a non-Abelian SU(2) ``magnetic'' field without dynamics \cite{Frohlich1992}. This effective ``magnetic'' field can be quite strong \cite{Nikolic2011a} and responsible for topologically protected edge states in two-dimensional TIs \cite{Murakami2004, Kane2005, Bernevig2006, Konig2007} -- just as the ordinary U(1) magnetic field is responsible for edge states in quantum Hall systems. Here we examine another possible analogy between U(1) and SU(2) fields: can a two-dimensional superconductor subjected to the Rashba spin-orbit coupling exhibit a vortex lattice like type-II superconductors in magnetic fields?

Spin-orbit-coupled superfluids have been studied in the context of cold atoms, motivated by the experimental progress in subjecting trapped two-component bosonic atoms to artificial gauge fields \cite{Spielman2009, Lin2011, Wang2012b, Cheuk2012}. A genuine Rashba spin-orbit coupling has not been created yet, but new technologies based on atom chips could realize it in the foreseeable future. Theoretical studies have mostly scrutinized the Rashba spin-orbit-coupled condensates of bosons that can move freely through the trap. Mean-field calculations favor ``plane wave'' and ``standing wave'' condensates formed at finite wavevectors \cite{Ho2011, Wang2010b, Zhai2012, Ozawa2012}, but quantum fluctuations may give rise to more unconventional states, perhaps even with fractional quasiparticles \cite{Sedrakyan2012}.

In this paper, we consider a two-dimensional lattice model of attractively interacting fermions that can form a paired spin-triplet superfluid in the presence of a Rashba spin-orbit coupling. The model describes a two-dimensional TI placed in contact with a conventional superconductor \cite{Nikolic2011a, Nikolic2012b}, but it can be reformulated to describe triplet exciton pairing among repulsively interacting electrons. The surface of a Kondo TI SmB$_6$ can also exhibit instabilities \cite{Nikolic2014b} that are addressed by this model. The choice of a lattice rather than a continuum model makes vortex arrays more competitive to the ``plane wave'' and ``standing wave'' states at the mean-field level. Indeed, we find condensates with a vortex lattice in a broad parameter regime when vortex cores of small size reside inside lattice plaquettes. The presented results follow from a numerical mean-field calculation at zero temperature. Quantum fluctuations enhanced by the low spatial dimensionality can give rise to vortex liquids.

The significance or our results is two-fold. First, the discovered vortex lattice is a novel state of matter -- an Abrikosov-type array of SU(2) topological defects that features spin-supercurrent loops and respects the time-reversal (TR) symmetry. The present study demonstrates a realistic microscopic model where such a vortex state can be stable. This is a different yet remarkably similar vortex array to the one identified by the Landau-Ginzburg analysis \cite{Nikolic2014} as a prominent metastable state in the continuum limit. Related but different states have been seen in a few studies of two-component interacting particles with spin-orbit coupling \cite{Cole2012, Radic2012, Cai2012}.

Second, the quantum melting of a vortex lattice can produce an incompressible quantum liquid \cite{nikolic:144507} whose excitations have fractional exchange statistics. This was numerically demonstrated in the context of rotating bosonic atoms, where the quantum melting of a U(1) Abrikosov lattice yields a fractional quantum Hall state \cite{Wilkin2000, Cooper2001, Regnault2003, Chang2005, Cooper2008}. Similar physics is expected in the present SU(2) system \cite{Nikolic2011a, Nikolic2011, Nikolic2012}. Both cold atom gases and quantum well heterostructures enable experimental manipulation of the quantum state. Destabilizing an existing vortex lattice at low temperatures by almost any means would allow experimental access to a fractional quantum liquid \cite{Nikolic2011a}. The non-Abelian fractional states created by the SU(2) Rashba flux could be useful for quantum computing. This work identifies the precise nature of the SU(2) vortices, which informs about the fractional statistics of quasiparticles in the corresponding vortex liquids \cite{Nikolic2012}.

The paper is organized as follows. Section \ref{secA} introduces the model, discusses its features and relates it to realistic physical systems. Section \ref{secB} presents the main results -- the mean-field phase diagram of SU(2) vortex states and other competing orders. Section \ref{secC} qualitatively analyzes fluctuation effects, and explains why incompressible quantum liquids with fractional excitations should also exist in the phase diagram of this model. The main discussion ends with conclusions in Section \ref{secConcl}. Certain special subjects and various technical details are discussed in Appendices.

\section{Competing superconducting orders by Rashba spin-orbit coupling}

\subsection{Model}\label{secA}

The model we study is given by the imaginary-time path integral 
\begin{equation}\label{Model}
\int \mathcal{D}\psi \mathcal{D}\eta \, \exp\left\lbrack -\int d\widetilde{t}
    \left( \psi^\dagger\frac{\partial\psi}{\partial{\widetilde{t}}} + H_0 + H_{\textrm{int}} \right) \right\rbrack \ ,
\end{equation}
expressed in terms of the four-component fermion (Grassmann) spinors $\psi$ and three-component bosonic (complex) spinors $\eta$. These fields live on a square lattice whose sites are labeled by $i$. The fermions $\psi_{i\sigma\tau}$ have spin $\sigma \in \lbrace \uparrow, \downarrow \rbrace$, and another two-state degree of freedom $\tau = \pm$. The latter is motivated by the Hamiltonian representation of an ultra-thin TI film \cite{Nikolic2011a}. A thick TI slab hosts protected helical Dirac-metal states on its top ($+$) and bottom ($-$) surfaces \cite{Fu2007}, with opposite helicity of spin-momentum locking. When these surfaces are brought close together in a film, the Dirac spectrum acquires a gap through inter-surface tunneling ($\Delta$), but the low-energy quasiparticles retain the $\tau$ degree of freedom associated with their surface belonging. The non-interacting Hamiltonian
\begin{equation}\label{H0}
H_{0} = \sum_{i} \left\lbrack
  -t\sum_{j\in i}\psi_{i}^{\dagger}\, e^{-i\tau^{z}\mathcal{A}_{ij}}\,\psi_{j}^{\phantom{\dagger}}
  +\psi_{i}^{\dagger}\left(-\mu+\Delta\tau^{x}\right)\psi_{i}^{\phantom{\dagger}} \right\rbrack
\end{equation}
describes fermion hopping on the tight-binding square lattice, in the presence of a static SU(2) gauge field defined on lattice links:
\begin{equation}\label{GaugeField}
\mathcal{A}_{ij}=-\mathcal{A}_{ji}\quad,\quad
  \mathcal{A}_{i,i+\hat{{\bf x}}}=a\sigma^{y}\quad,\quad\mathcal{A}_{i,i+\hat{{\bf y}}}=-a\sigma^{x} \ .
\end{equation}
This gauge field, written in terms of the Pauli matrices $\sigma^{x,y,z}$ that couple to spin, implements the Rashba spin-orbit coupling of strength $a$. The Pauli matrices $\tau^{x,y,z}$ operate on the ``surface index'' $\tau$. The fermions' SU(2) charge $\tau^z$ depends on the surface they visit, and the $\Delta\tau^x$ term directly describes inter-surface tunneling that creates a bandgap. The fermion density is controlled by the chemical potential $\mu$ in our formalism.

Interestingly, the 100 crystal surface of a Kondo TI SmB$_6$ has similar features. It hosts three helical Fermi pockets in the vicinity of $\Gamma$, X and Y points of the surface's 1$^\textrm{st}$ Brillouin zone \cite{Nickerson1971, Farberovich1983, Takimoto2011, Kang2013, Lu2013b}. The $\Gamma$ pocket has the opposite helicity of spin-momentum locking than portions of the X and Y pockets \cite{Nikolic2014b}. Collective excitations seen by neutron scattering in SmB$_6$ are precisely poised to couple these pockets in either Cooper or exciton channels \cite{Fuhrman2014, Nikolic2014c}. Any coupling between the pockets of opposite helicity can lead to triplet pairing of the kind we investigate in this paper.

The SU(2) flux $\Phi$ of (\ref{GaugeField}) on a square plaquette, with corners $1,2,3,4$ enumerated in a clockwise sense, can be defined as a matrix by:
\begin{equation}
e^{i\tau^z\Phi} = e^{i\tau^{z}\mathcal{A}_{12}} e^{i\tau^{z}\mathcal{A}_{23}} e^{i\tau^{z}\mathcal{A}_{34}} e^{i\tau^{z}\mathcal{A}_{41}} \ .
\end{equation}
Depending on which of the four corners is labeled by $1$, one finds
\begin{eqnarray}\label{Flux}
\Phi \!\!&=&\!\! \frac{\arccos\Bigl(1-2\sin^{4}(a)\Bigr)}{\sqrt{1+2\tan^{2}(a)}}\tau^{z}
  \Bigl\lbrack\sigma^{z}+\tan(a\tau^{z})(\pm\sigma^{x}\pm\sigma^{y})\Bigr\rbrack \nonumber \\
\!\!&=&\!\! 2a^{2}\tau^z\sigma^{z}+2a^{3}(\pm\sigma^{x}\pm\sigma^{y})+\mathcal{O}(a^{4})
\end{eqnarray}
with four different combinations of $\pm$ signs. The leading order term $\Phi \to 2a^{2}\tau^z\sigma^{z}$ in the limit of small $a$ is finite and independent of the exact path around a plaquette; this becomes the ``magnetic'' Yang-Mills flux
\begin{equation}
\Phi \to \hat{\bf z} \Bigl( \boldsymbol{\nabla}\times\boldsymbol{\mathcal{A}}
  -i\tau^z \boldsymbol{\mathcal{A}}\times\boldsymbol{\mathcal{A}} \Bigr)
\end{equation}
in the equivalent continuum-limit single-particle Hamiltonian:
\begin{eqnarray}
H_0 &\to& \frac{1}{2m} ({\bf p} - \tau^z \boldsymbol{\mathcal{A}})^2 -\mu' + \Delta\tau^{x}  \\
&=& \frac{p^2}{2m} +v \tau^z \hat{\bf z} (\boldsymbol{\sigma} \times {\bf p}) -\mu' + \Delta\tau^{x} \nonumber \ .
\end{eqnarray}
Here, $t=1/2m$, $\mu'=\mu+4t$, $v=a/m$, $\boldsymbol{\mathcal{A}} \to -mv (\hat{\bf z} \times \boldsymbol{\sigma})$ and $\boldsymbol{\sigma} = (\sigma^x, \sigma^y, \sigma^z)$, using the units in which $\hbar$ and the lattice constant are equal to $1$. The eigenvalues of the matrix $\Phi$ are SU(2) gauge-invariant and their finite values lead to an Aharonov-Bohm effect just like the analogous U(1) magnetic field, even though $\textrm{tr}(\Phi) = 0$ due to the TR symmetry.


Interactions among the fermions are generated by a Hubbard-Stratonovich field, the $S=1$ bosonic spinor $\eta$:
\begin{eqnarray}\label{Hint}
H_{\textrm{int}} \!\!&=&\!\! \sum_{i}\Biggl\lbrack
    U_t\left(|\eta_{i\uparrow}|^2+|\eta_{i\downarrow}|^2\right)+U_{t0}|\eta_{i0}|^2 \\
&&  +\eta_{i\uparrow}^{*}\psi_{i\uparrow+}^{\phantom{*}}\psi_{i\uparrow-}^{\phantom{*}}
    +\eta_{i\downarrow}^{*}\psi_{i\downarrow+}^{\phantom{*}}\psi_{i\downarrow-}^{\phantom{*}} \nonumber \\
&&  +\eta_{i0}^{*}\frac{\psi_{i\uparrow+}^{\phantom{*}}\psi_{i\downarrow-}^{\phantom{*}}
      +\psi_{i\downarrow+}^{\phantom{*}}\psi_{i\uparrow-}^{\phantom{*}}}{\sqrt{2}}+h.c.\Biggr\rbrack \ . \nonumber
\end{eqnarray}
This is the part of the most general decoupling of short range attraction that leads to Cooper pairing in spin-triplet channels \cite{Nikolic2011a}. The remaining part, responsible for singlet pairing, is discussed in the Appendix and otherwise neglected -- singlets are completely suppressed at the mean-field level in the portion of the phase diagram that we analyze, and, more importantly, they are impaired by spin-orbit SU(2) neutrality. Triplet Cooper pairs are able to gain energy from the Rashba spin-orbit coupling in proportion to the momentum they carry \cite{Nikolic2012a}.

The model (\ref{Model}) is invariant under TR and has the full point-group symmetry of the square lattice. The only continuous symmetry of the model for generic $a$ is the U(1) symmetry responsible for charge conservation. This symmetry is spontaneously broken in a superfluid state. However, the continuous symmetry group is larger if $a = n \frac{\pi}{2}$, $n \in \mathbb{Z}$. At $a=0$, the spin ``up'' and ``down'' states are decoupled, so one can apply an independent global U(1) transformation on each spin projection. At $a=\pi/2$, the nearest-neighbor hopping terms in (\ref{H0}) reduce to $\propto \psi_i^\dagger \sigma^{x,y} \psi_j^{\phantom{\dagger}}$, which allows the following sublattice-dependent symmetry transformations on a bipartite lattice:
\begin{eqnarray}\label{Sym2}
\eta_{A\uparrow} \to e^{i\theta_A}\eta_{A\uparrow} \quad&,&\quad
  \eta_{B\uparrow} \to e^{i\theta_B}\eta_{B\uparrow} \\
\eta_{A\downarrow} \to e^{i\theta_B}\eta_{A\downarrow} \quad&,&\quad
  \eta_{B\downarrow} \to e^{i\theta_A}\eta_{B\downarrow} \nonumber \\
\eta_{A0} \to e^{i\frac{\theta_A+\theta_B}{2}}\eta_{A0} \quad&,&\quad
  \eta_{B0} \to e^{i\frac{\theta_A+\theta_B}{2}}\eta_{B0} \nonumber \\
\psi_{A\uparrow\pm} \to e^{i\frac{\theta_A}{2}}\psi_{A\uparrow\pm} \quad&,&\quad
  \psi_{B\uparrow\pm} \to e^{i\frac{\theta_B}{2}}\psi_{B\uparrow\pm} \nonumber \\
\psi_{A\downarrow\pm} \to e^{i\frac{\theta_B}{2}}\psi_{A\downarrow\pm} \quad&,&\quad
  \psi_{B\downarrow\pm} \to e^{i\frac{\theta_A}{2}}\psi_{B\downarrow\pm} \nonumber
\end{eqnarray}
This is a double global U(1) transformation specified by two angles $\theta_A$ and $\theta_B$, where $A,B$ indicate the lattice sites belonging to one or the other sublattice. All properties of the model are periodic under $a \to a+2\pi$, while the change $a \to a+\pi$ is equivalent to changing the sign of $t$. The spectrum is invariant under $a \to -a$, since this corresponds to the lattice inversion through a point (without affecting spin).

\subsection{The mean-field phase diagram}\label{secB}

In this paper we analyze the zero-temperature phase diagram of the model (\ref{Model}), beginning with the mean-field approximation. We consider various spatially dependent but static order parameters $\eta = (\eta_\uparrow, \eta_0, \eta_\downarrow)$, and search for the one that minimizes the ground state energy at any given values of $\mu, \Delta, a, U_t, U_{t0}$, etc \cite{ARGO}. This is equivalent to the usual self-consistent solution of the Bogoliubov-de Gennes equations for a superconductor. We assume that the order parameter $\eta$ belongs to one of these categories:
\begin{enumerate}
\item spatially periodic with a unit-cell of $l_x \times l_y$ sites: freely vary $\eta_i$ at every site of the unit-cell;
\item commensurate pair-density wave (PDW) at a wavevector ${\bf Q} = (q_x,q_y)$: vary ${\bf Q}$ and the Fourier amplitudes of $\eta$;
\item incommensurate plane wave state (PWS) at a wavevector ${\bf Q} = (q_x,q_y)$: vary ${\bf Q}$ and the amplitude of $\eta$.
\end{enumerate}
The first category is the most general one among all periodic states, but computationally demanding if the unit-cell is large. For example, we need 12 complex numbers to specify a periodic state with a $l_x = l_y = 2$ unit-cell. The second category is a subset of the first category, but faster to compute. The PDW order parameter is restricted to the form:
\begin{equation}\label{PDW}
\eta_i = \eta_+ e^{i{\bf Q}{\bf r}_i} + \eta_- e^{-i{\bf Q}{\bf r}_i} \ ,
\end{equation}
where ${\bf r}_i = (x_i,y_i)$ are the integer-valued coordinates of the lattice site $i$, and $\bf Q$ is a commensurate wavevector in the first Brillouin zone (i.e. there is an integer $n$ such that $n{\bf Q}$ is a reciprocal lattice vector). It takes 6 complex numbers to specify the two triplet amplitudes $\eta_\pm$, and another two rational numbers to specify $\bf Q$. Finally, the third category captures the plane wave subset of all category-2 states:
\begin{equation}\label{PWS}
\eta_i = \eta_{\textrm{pw}} e^{i{\bf Q}{\bf r}_i} \ .
\end{equation}
It takes 3 complex numbers to specify the amplitude $\eta_{\textrm{pw}}$, and two real numbers to specify $\bf Q$. A plane-wave state can be handled even when it is incommensurate, since it respects the lattice translation symmetries up to a global gauge transformation. Within each of these categories, we construct the Bogoliubov-de Gennes Hamiltonian and diagonalize it to determine the superfluid mean-field ground state. Then, we vary all the quantities that specify the order parameter in its category until we find a local minimum of the ground state energy at any point in the phase diagram. At the end, the absolute ground state energies in all categories are compared to identify the best order parameter. A more technical discussion of this procedure and its accuracy can be found in the Appendix.

\begin{figure*}
\includegraphics[height=2.2in]{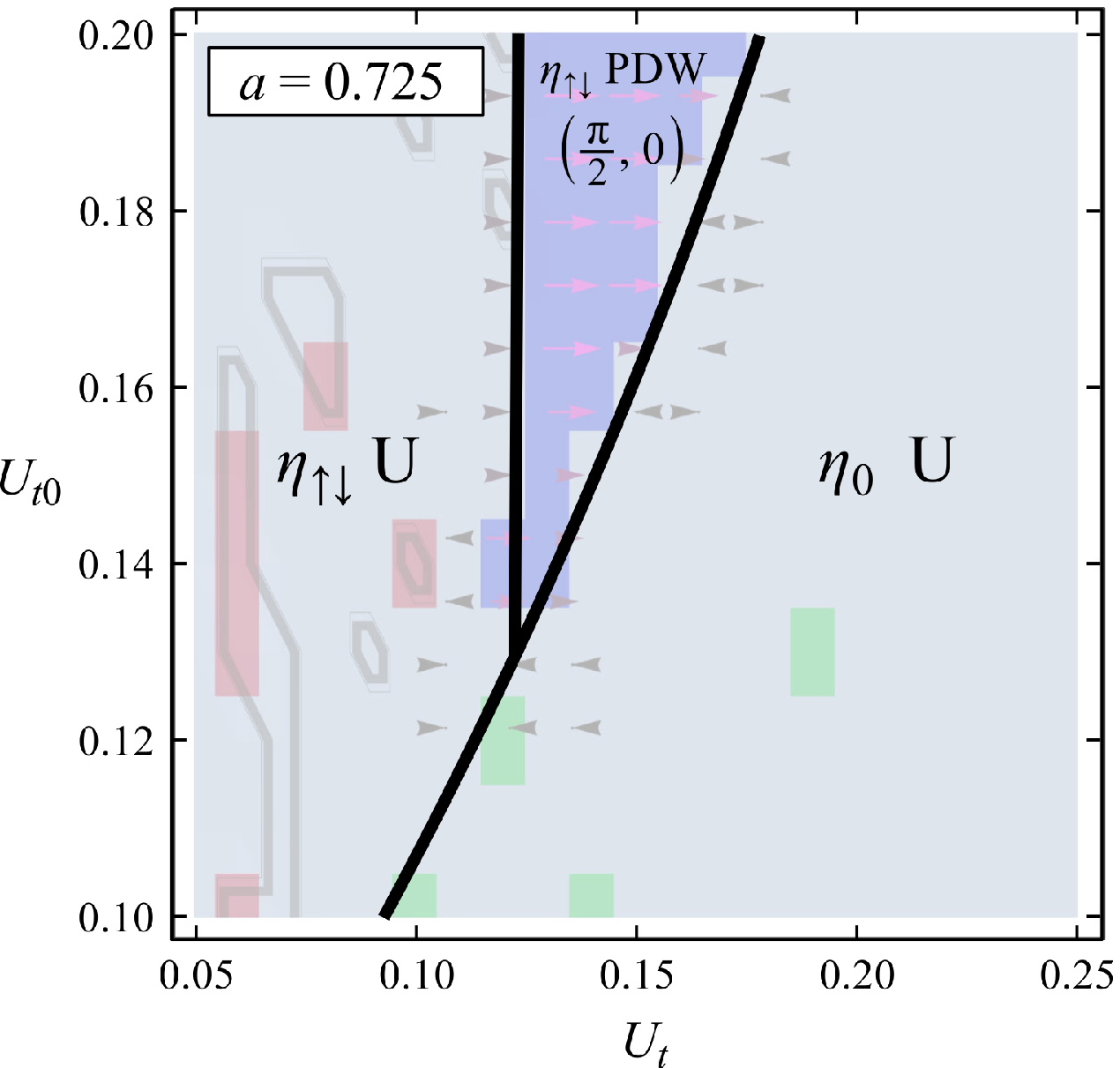} \hskip 0.03in
\includegraphics[height=2.2in]{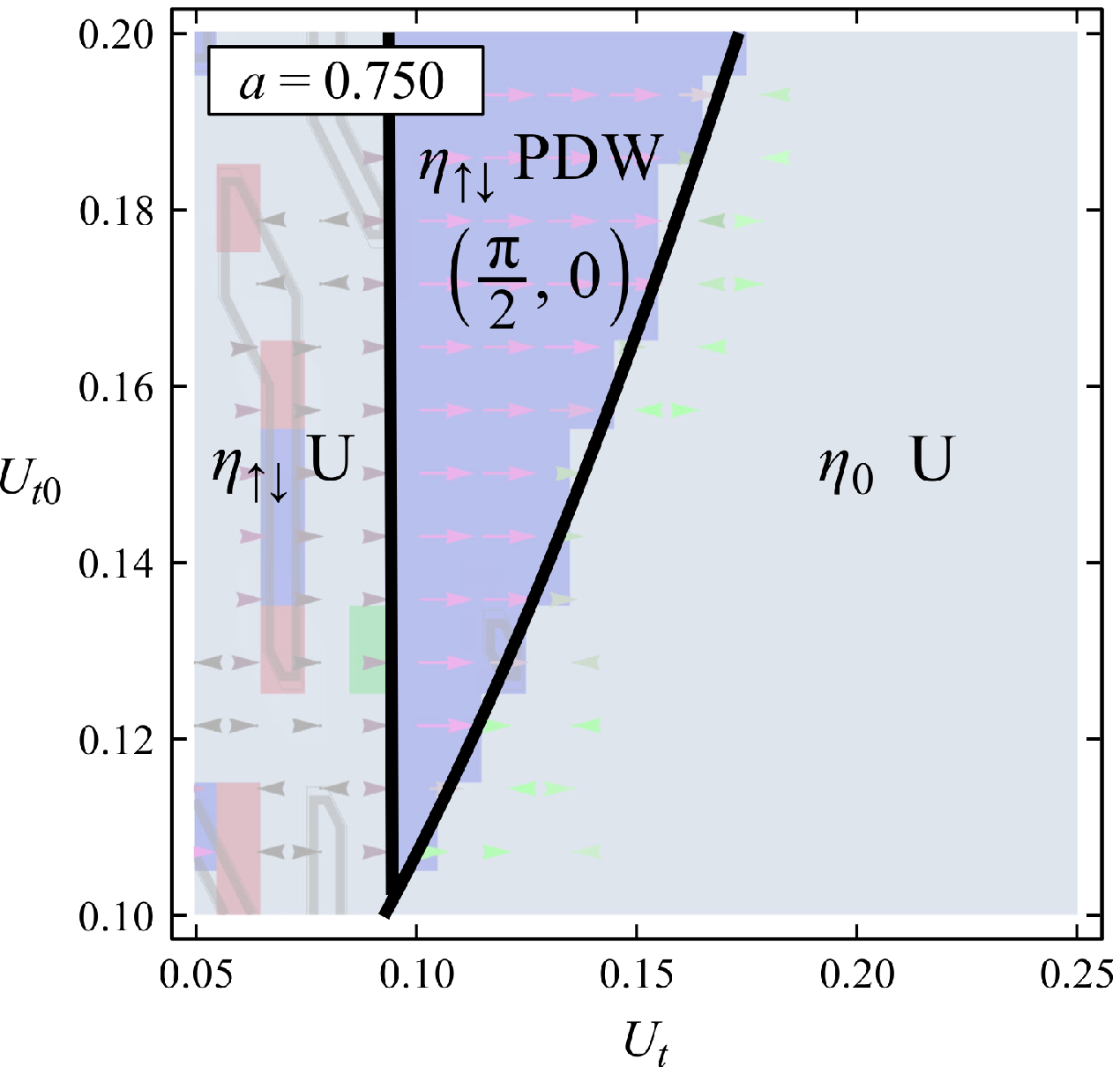} \hskip 0.03in
\includegraphics[height=2.2in]{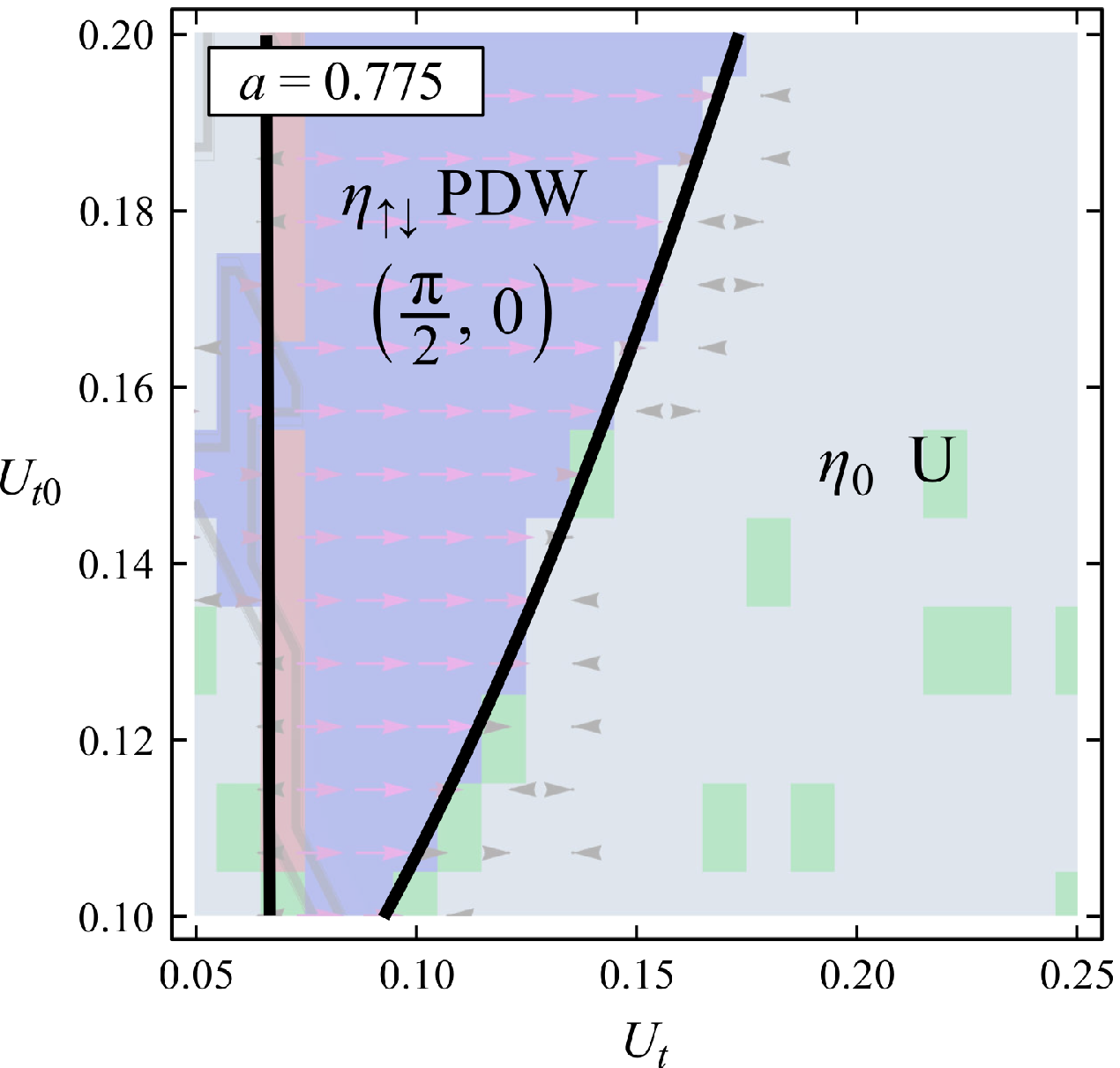}
\includegraphics[height=2.2in]{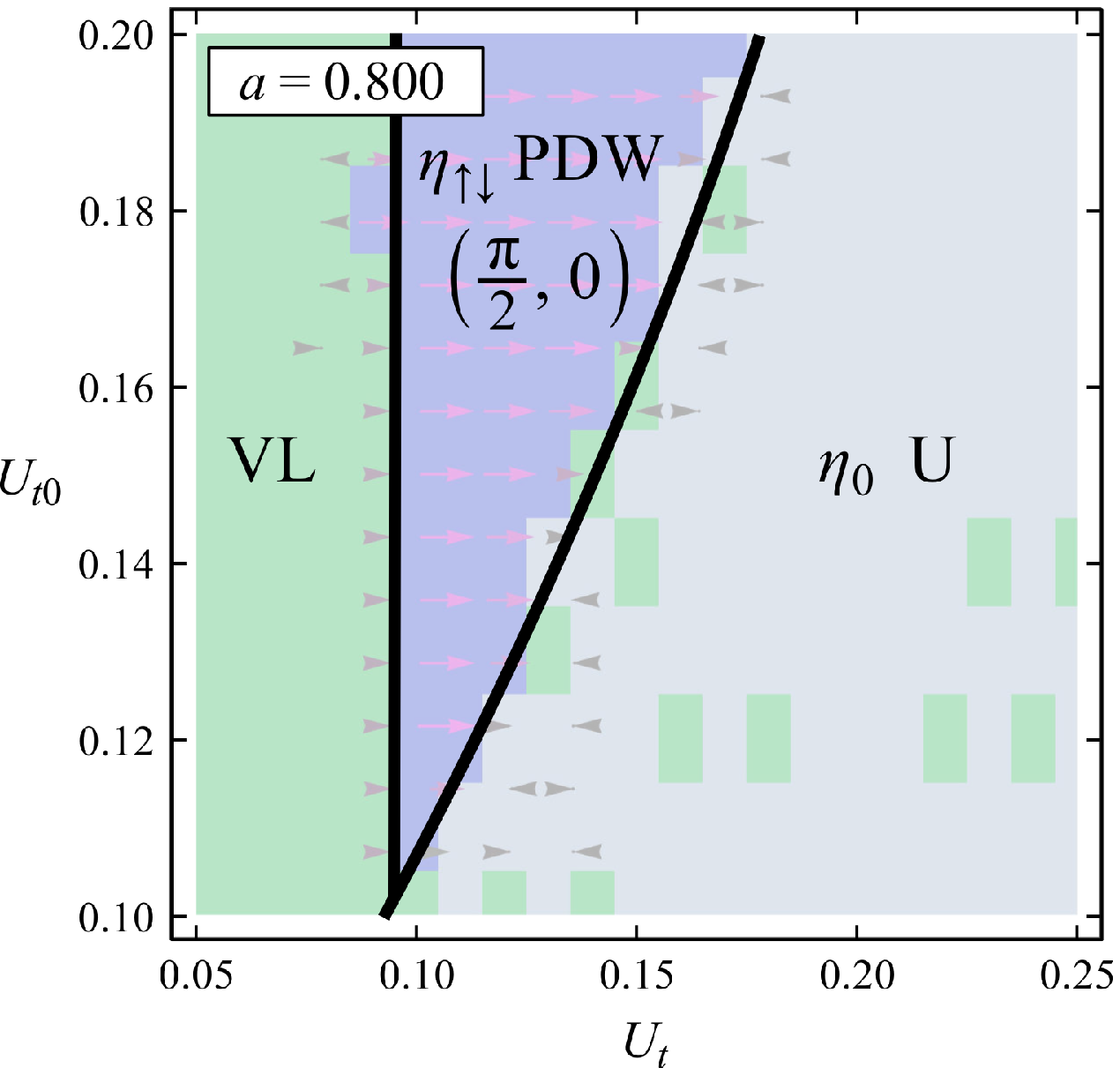} \hskip 0.03in
\includegraphics[height=2.2in]{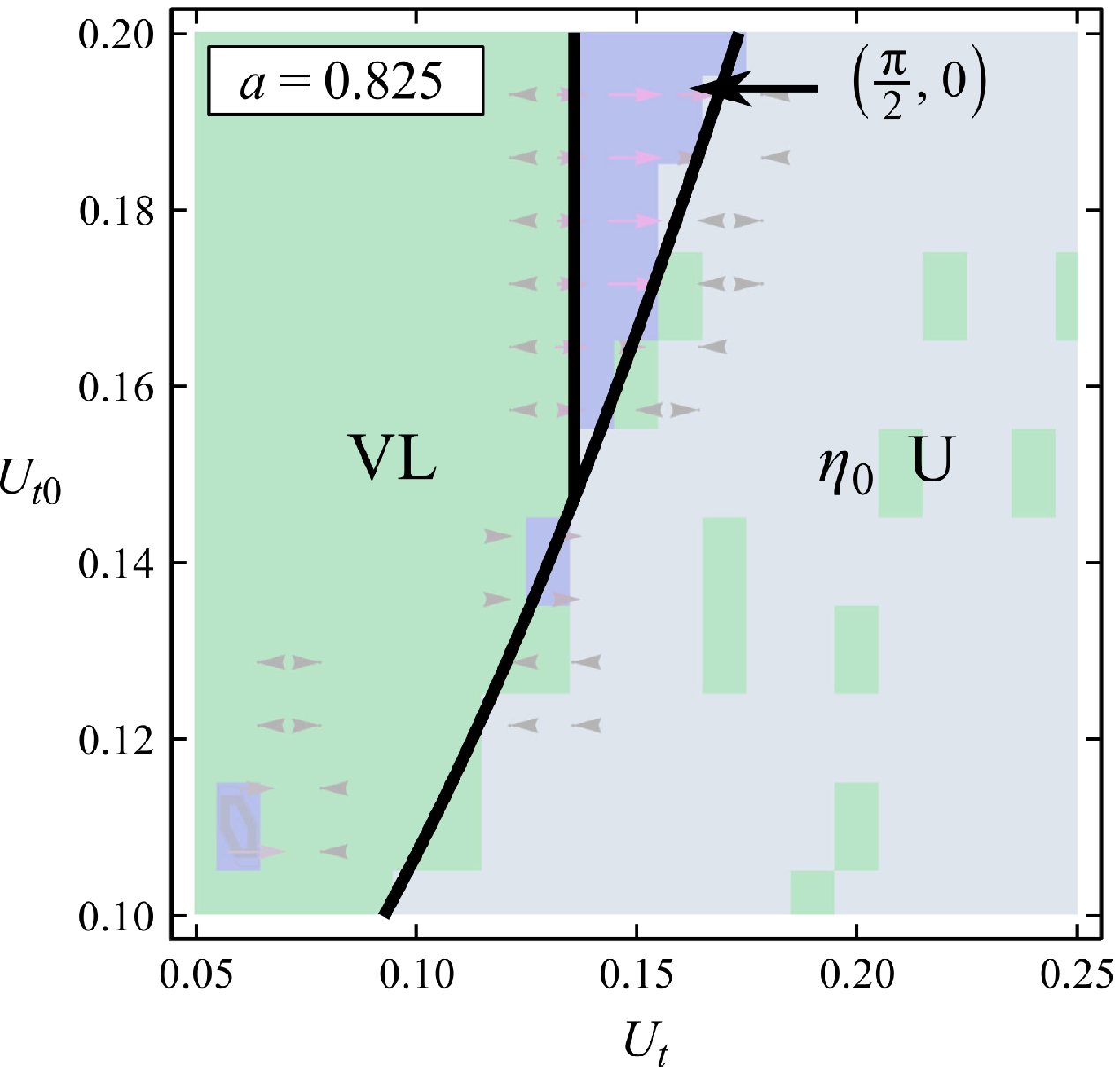} \hskip 0.03in
\includegraphics[height=2.2in]{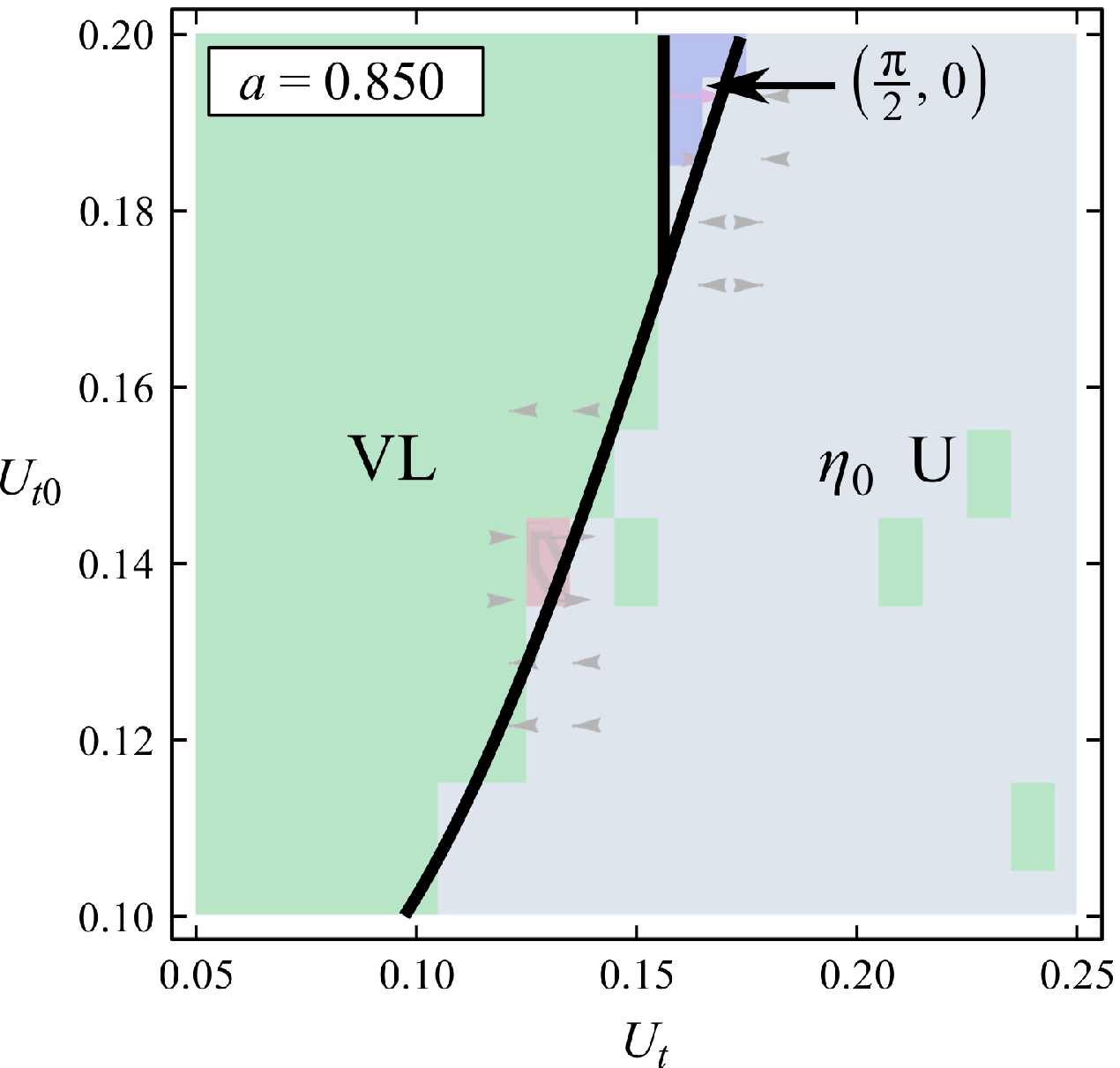}
\includegraphics[height=2.2in]{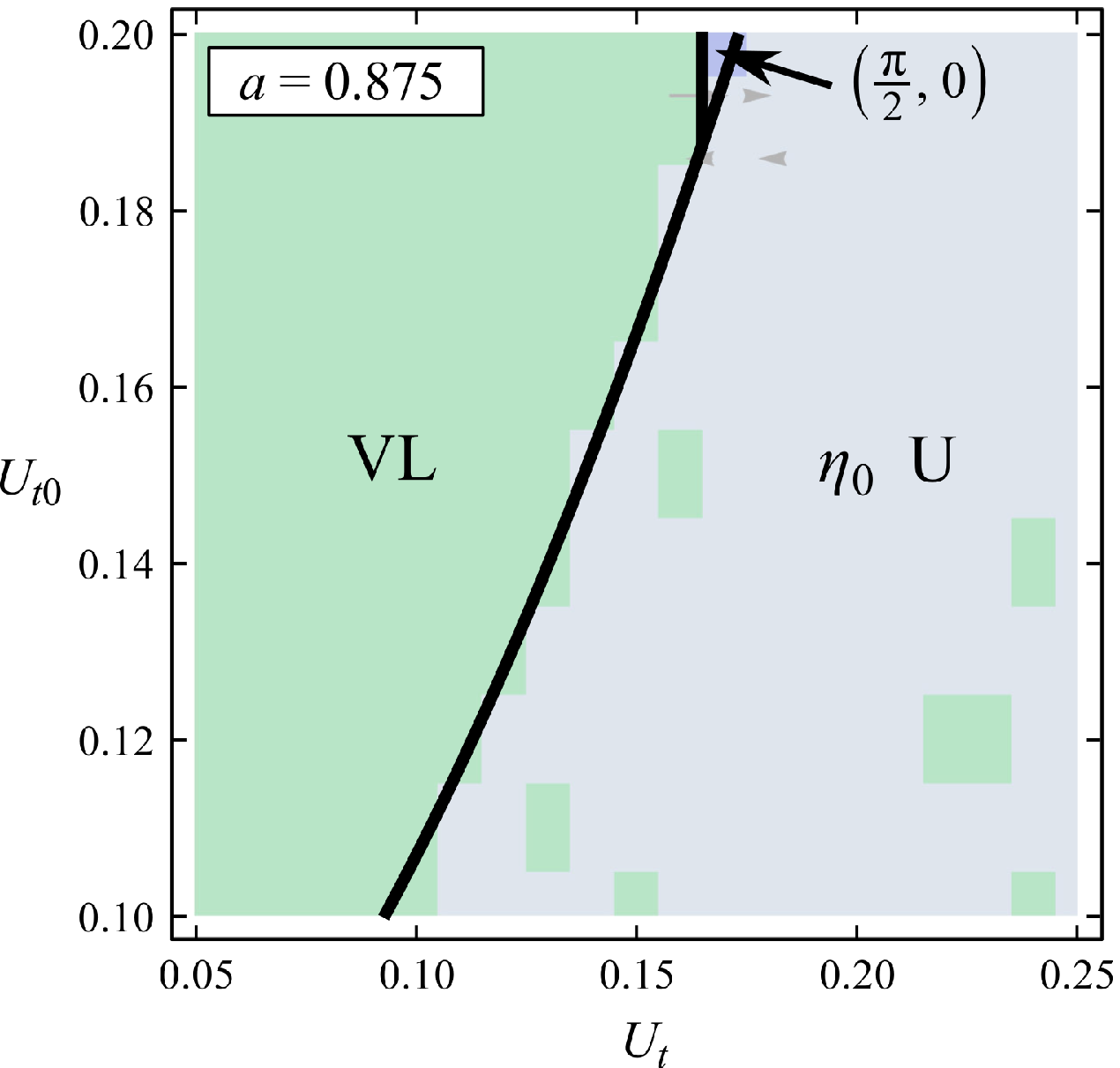} \hskip 0.03in
\includegraphics[height=2.2in]{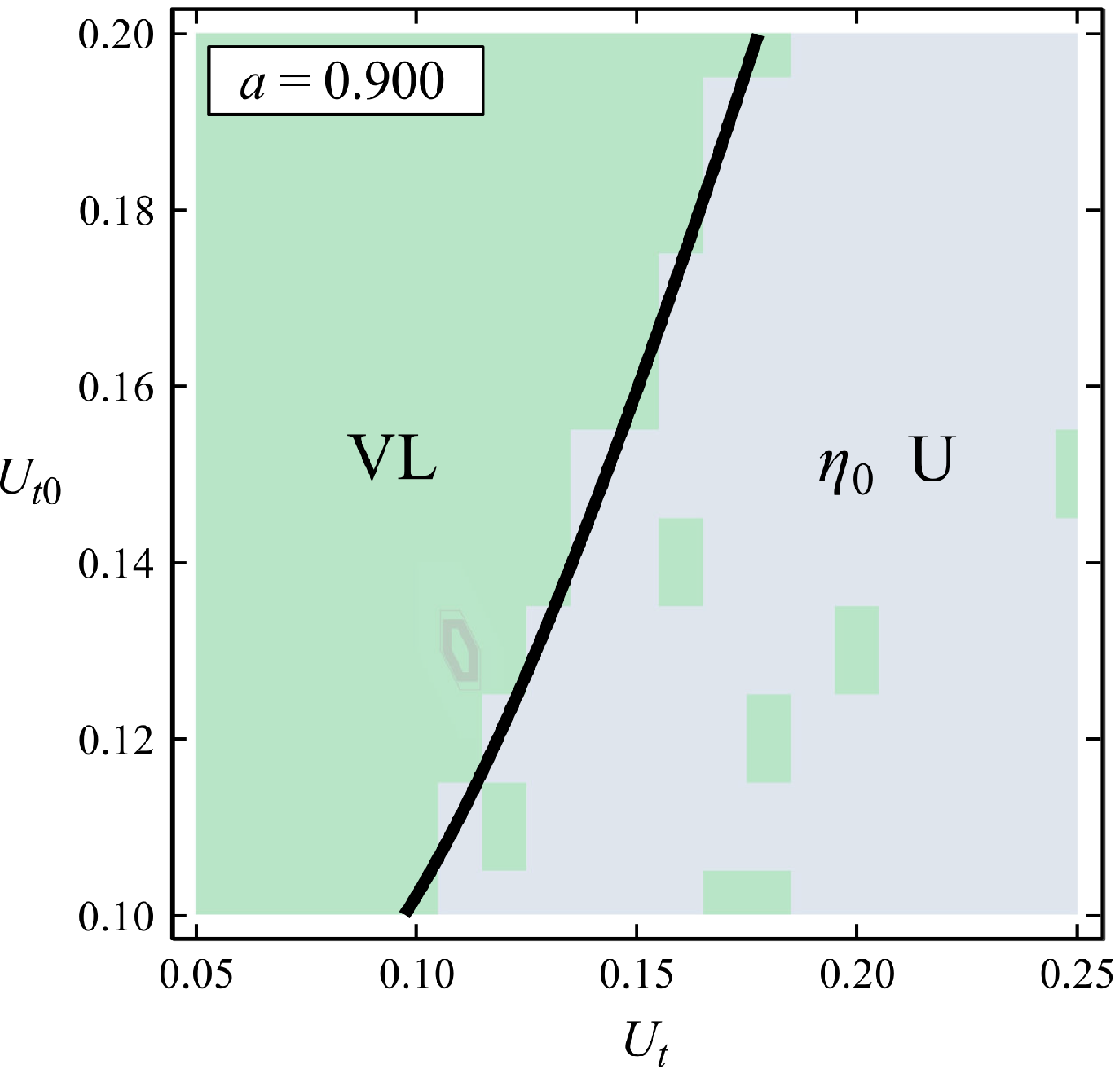} \hskip 0.03in
\includegraphics[height=2.2in]{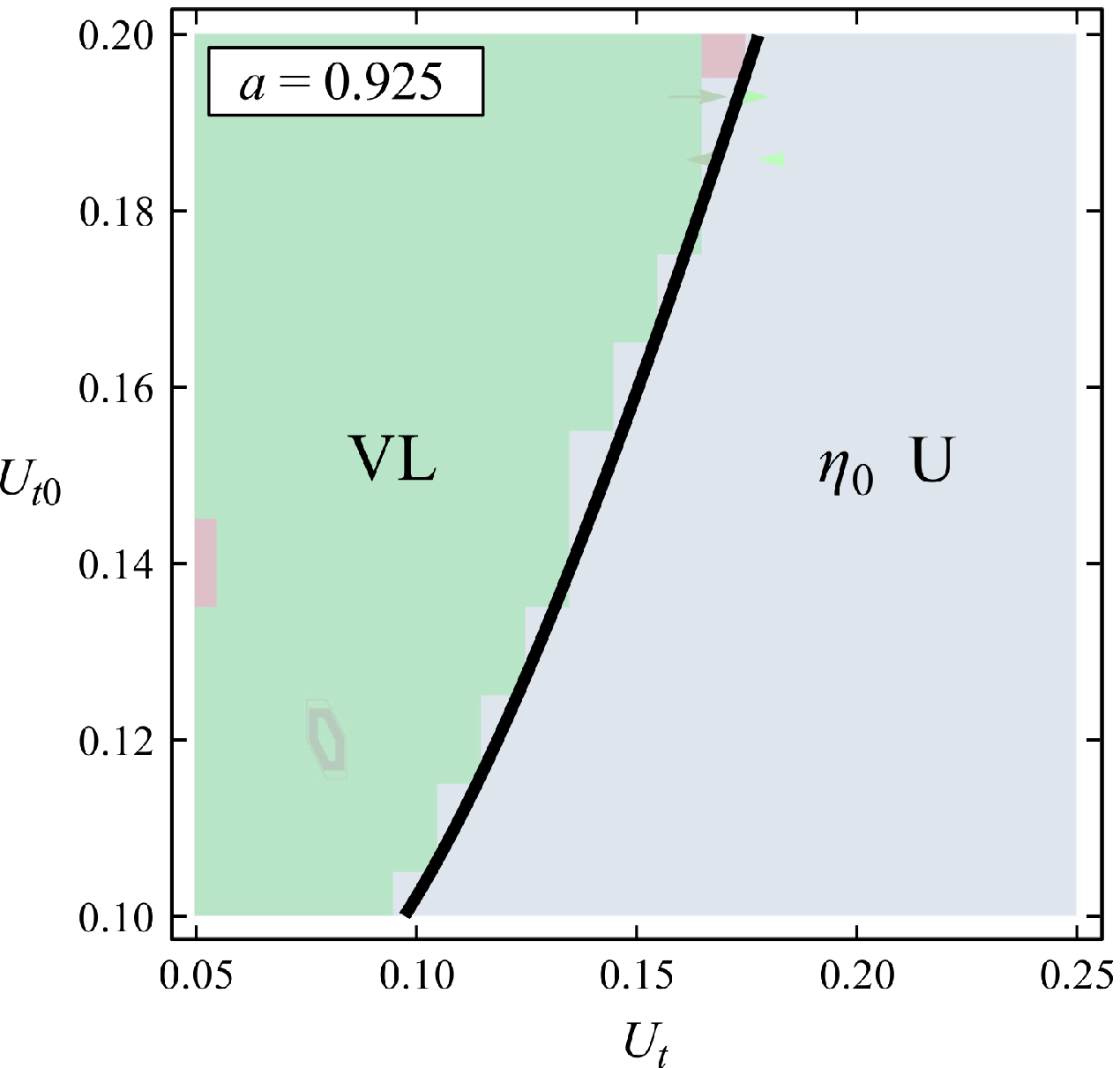}
\caption{\label{phdiag1}The mean-field phase diagram of paired states in the region that features pair density waves (PDW) and vortex lattices (VL). The phase transitions (black lines) are generally 1st order, and all found states are TR-invariant. The denoted phase labels and transitions are guides for the eye. The raw data is color-coded in the background (see Appendix for explanation). The order parameter is dominated by the spinful components $\eta_\uparrow, \eta_\downarrow$ for $U_t\lesssim U_{t0}$ and the ``symmetric triplet'' component $\eta_0$ for $U_t\gtrsim U_{t0}$. The interaction couplings $U_t, U_{t0}$ are expressed in the units of $t^{-1}$ and $\mu=0, \Delta=t$, where $t$ is the hopping constant of the Hamiltonian (\ref{Model}). The order parameter magnitude in the same units of energy is $|\eta|\sim t$, implying that the results correspond to a strong coupling regime.}
\end{figure*}

The main results are illustrated in Fig.\ref{phdiag1}. This is the phase diagram of non-trivial superfluid states that take advantage of the spin-orbit coupling $a$, plotted as a function of the two interaction couplings $U_t, U_{t0}$. The chemical potential $\mu=0$ is kept in the middle of the bandgap created by $\Delta> 0$, but interactions are strong enough to cause pairing. The first striking feature is the existence of several competing orders that reflect the frustrated dynamics. Frustration is created by the competition between the lattice constant (equal to $1$ in our units), the SU(2) ``magnetic'' length ($\sim a^{-1}$), and the pairing coherence length. The discovered orders have very similar energies, and differ by no more than $1-2 \%$ of the absolute energy value in many scrutinized parts of the phase diagram. The achieved absolute error of any calculated energy is much smaller than the energy differences between the found states. 

The line $U_t \approx U_{t0}$ separates two different regimes. For $U_t \gtrsim U_{t0}$, the symmetric triplet component $\eta_0$ of the order parameter costs less energy than the spinful components $\eta_\uparrow$ and $\eta_\downarrow$. The ensuing $\eta_0$ condensates are found to be uniform and with completely suppressed $\eta_\uparrow = \eta_\downarrow = 0$. Conversely, the region $U_t \lesssim U_{t0}$ hosts condensates dominated by $\eta_\uparrow$ and $\eta_\downarrow$, typically with $\eta_0=0$. The phase transition between the two types of condensates is first order within numerical resolution. The vanishing of $\eta_0=0$ when $U_t \lesssim U_{t0}$ explains the tendency of the other phase boundaries to be vertical (the quasiparticles end up decoupled from $U_{t0}$ in the mean-field approximation, so the phase boundaries do not depend on $U_{t0}$).

The region $U_t \lesssim U_{t0}$ with $\eta_{\uparrow\downarrow} \neq 0$ hosts TR-invariant vortex lattice condensates at $a \ge 0.8$. The appearance of dense vortex lattices is related to the large SU(2) flux (\ref{Flux}), maximized at $a \approx 0.96$. A TR-invariant vortex lattice needs both $\eta_{\uparrow}$ and $\eta_{\downarrow}$ order parameter components to establish a pattern of spin currents that gain energy through the Rashba spin-orbit coupling. It is, then, natural that vortex states live only in the $U_t \lesssim U_{t0}$ regions.

\begin{figure}
\subfigure[{}]{\includegraphics[height=1.6in]{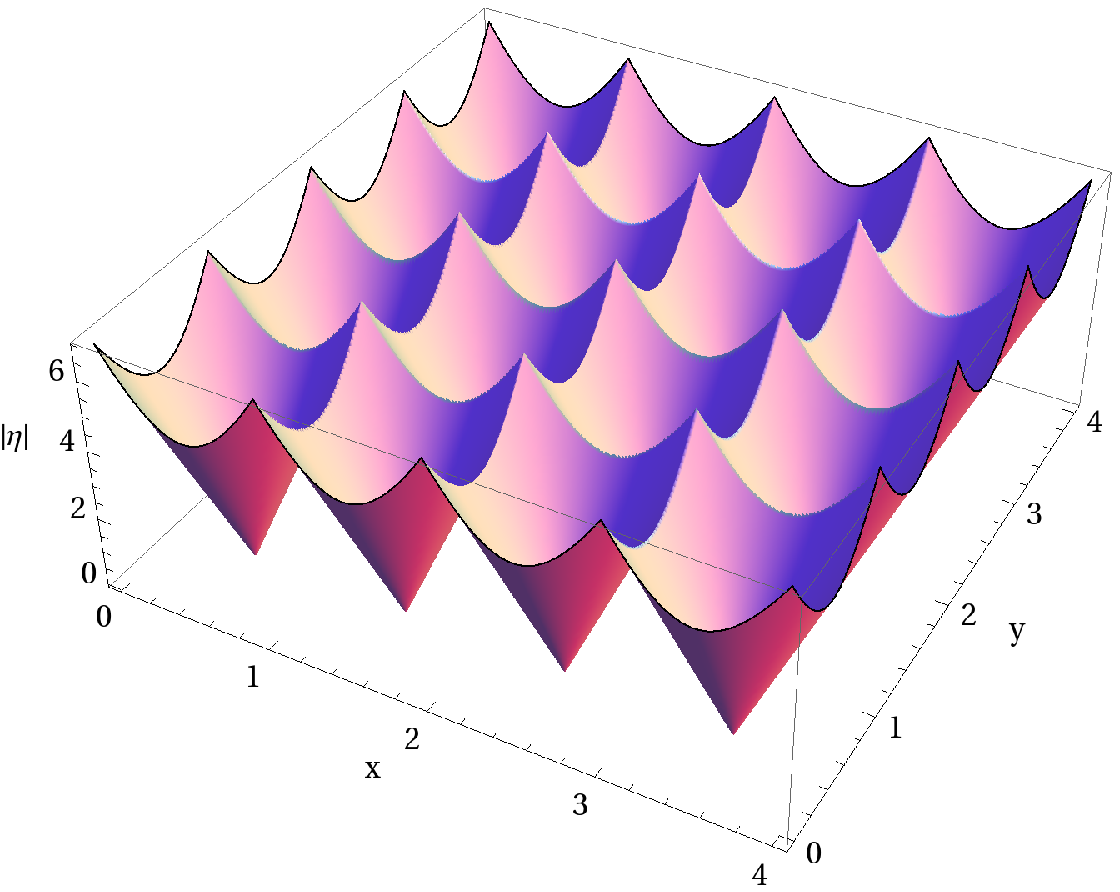}}
\subfigure[{}]{\includegraphics[height=1.2in]{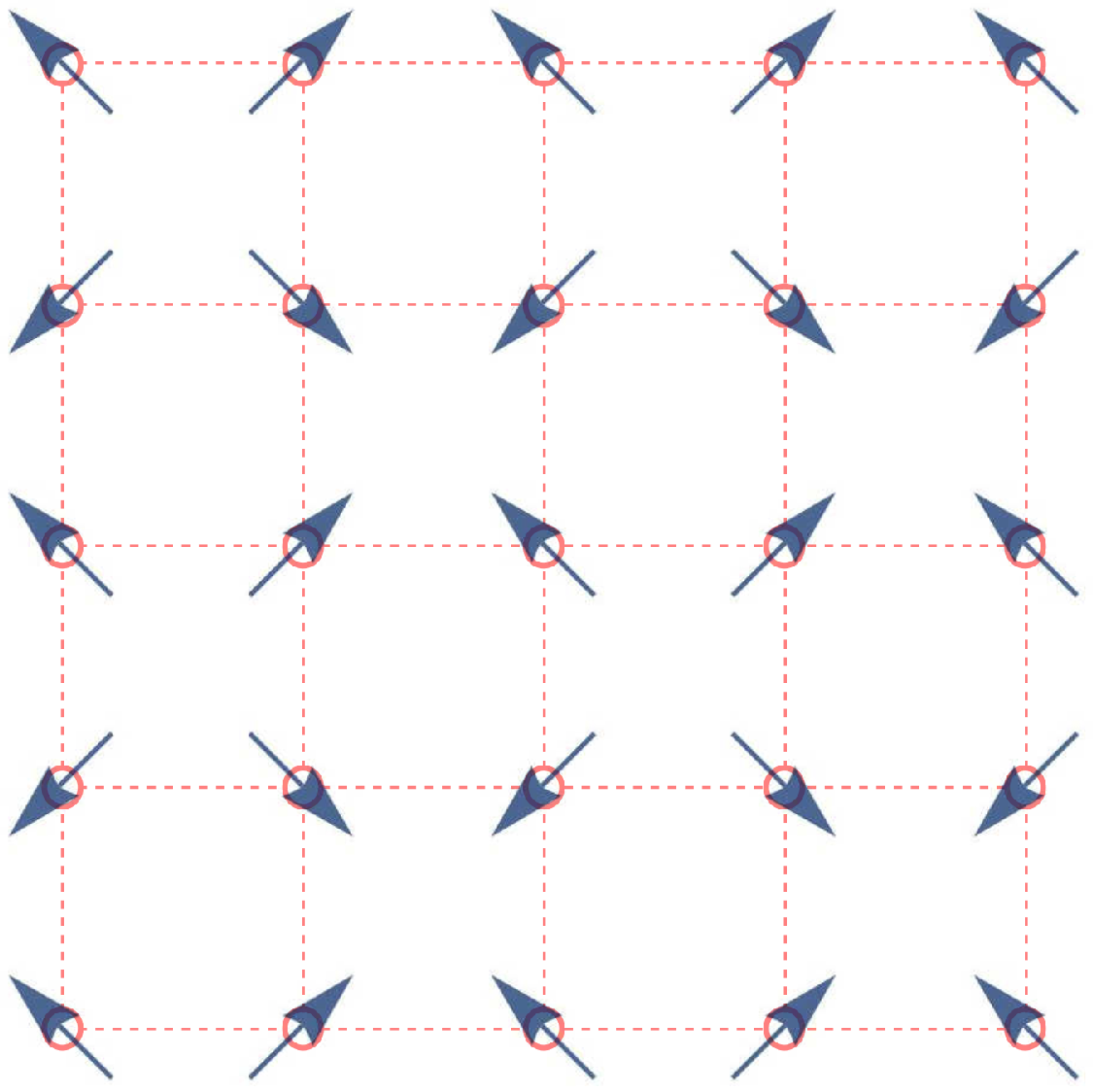}}
\subfigure[{}]{\includegraphics[height=1.4in]{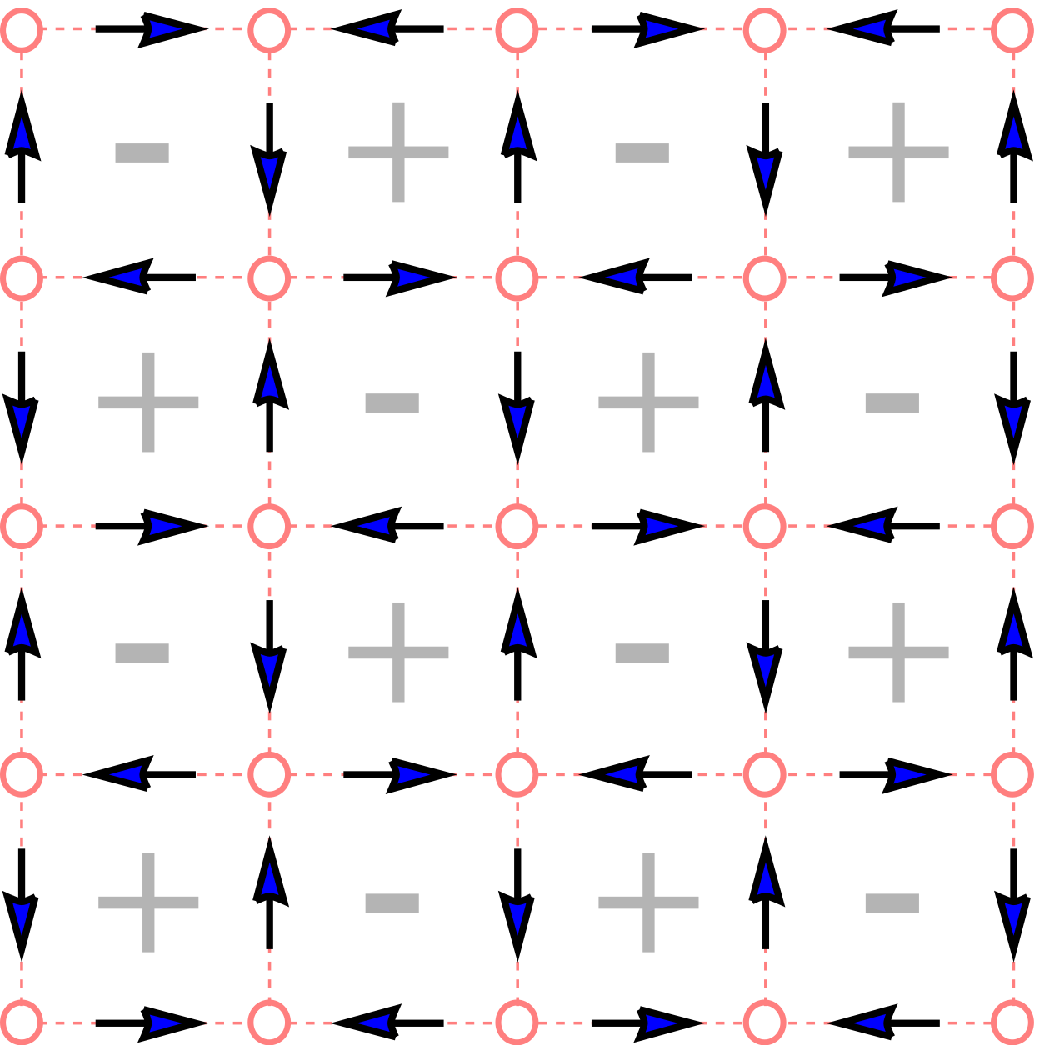}} \hskip 0.2in
\subfigure[{}]{\includegraphics[height=1.4in]{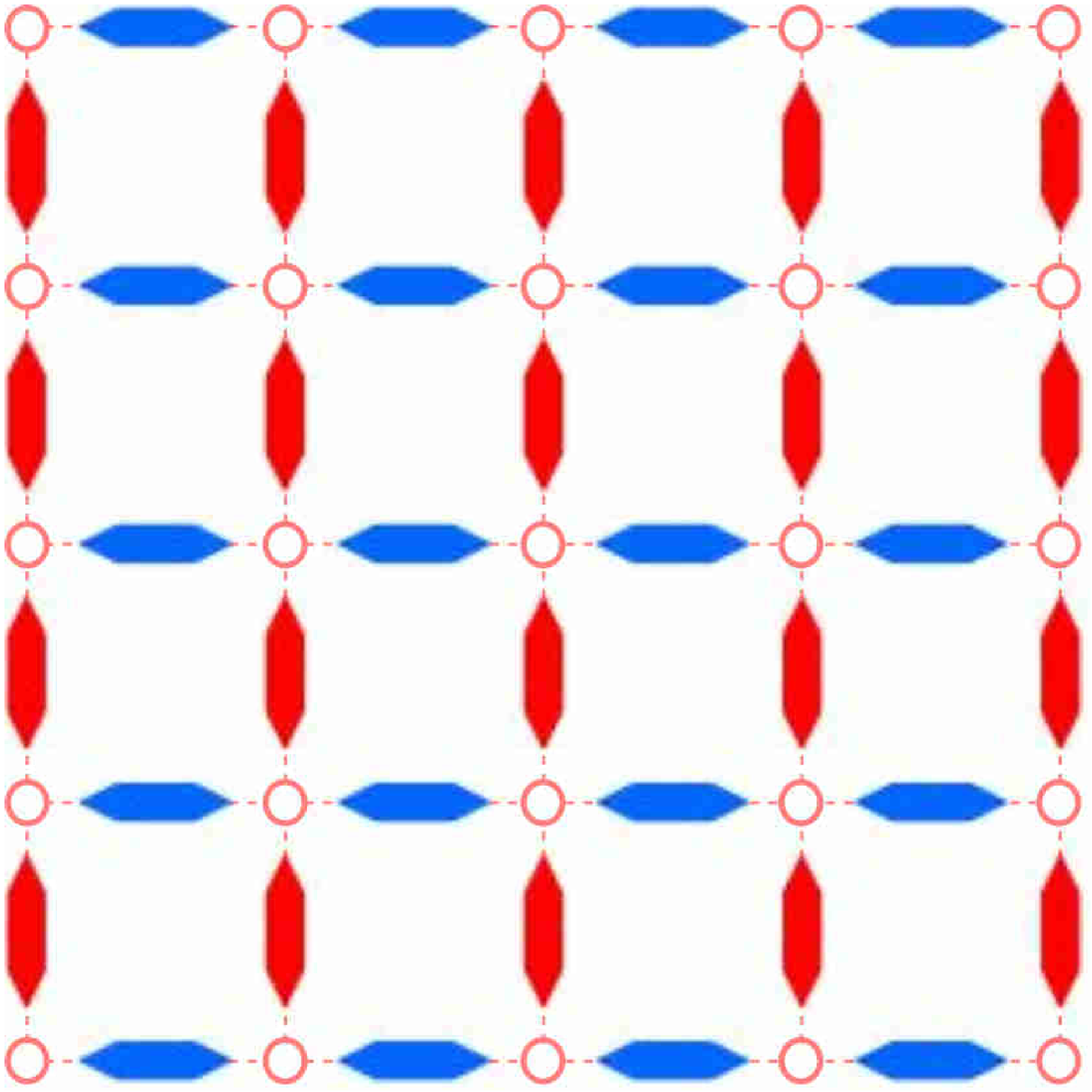}}
\caption{\label{vl-op}A typical order parameter $\eta = (\eta_\uparrow,\eta_0,\eta_\downarrow) \approx (\bar{\eta},0,\bar{\eta}^*)$ of the TR-invariant SU(2) vortex array in the $2\times2$-site unit-cell, obtained by category-1 calculations. (a) The magnitude $|\bar{\eta}|$ of the linearly-extrapolated $\bar{\eta}$ from the integer-valued lattice site coordinates to real-valued coordinates (in order to visualize singularities inside lattice plaquettes). (b) The phase $\theta$ in $\bar{\eta}=|\bar{\eta}|e^{i\theta}$. (c) The pattern of $J^z$ spin currents (\ref{SpinCurrents}) on lattice bonds exhibits a checkerboard array of vortices (+) and antivortices (-). (d) $K^{xx}$ defined by (\ref{BondScalars}), with the same magnitude on all bonds but different sign on horizontal and vertical bonds; red and blue denote $K>0$ and $K<0$ on a given bond respectively. Note that $K^{yy}$ is the same as $K^{xx}$ rotated by 90 degrees.}
\end{figure}

The vortex lattice states were found by the category-1 search using the minimal unit-cell of $2\times 2$ sites. Their order parameter, shown in Fig.\ref{vl-op}, is TR-invariant and features a periodic structure of spin currents on lattice links:
\begin{equation}\label{SpinCurrents}
J_{ij}^{k} = -\frac{i}{2}\Bigl\lbrack\eta_i^{*}S^{k}\eta_j^{\phantom{k}}-\eta_j^{*}S^{k}\eta_i^{\phantom{k}}\Bigr\rbrack \ ,
\end{equation}
where $S^k$ ($k\in\lbrace x,y,z \rbrace$) are spin projection matrices in the $S=1$ representation. Partial understanding of this state follows from the Landau-Ginzburg theory of two-dimensional $S=1$ superfluids with Rashba spin-orbit coupling \cite{Nikolic2014}, which predicts that similar checkerboard vortices of $J^z$ spin-currents are metastable. In the continuum Landau-Ginzburg picture, the spin-orbit energy is actually gained through ``helical'' spin currents $J^{x,y}$ -- the pattern of $J^{x,y}$ becomes singular and then drives the appearance of $J^z$ spin-current vortices. However, in the present calculation, $\eta_0=0$ renders $J^{x,y}=0$ in the vortex state, so the spin-orbit energy is gained by an intrinsically lattice mechanism. By gauge symmetry \cite{Nikolic2011a, Nikolic2012a}, the triplets are effectively coupled to the SU(2) gauge field:
\begin{eqnarray}\label{EffTheory}
H_{\textrm{eff}} &=& -t' \sum_{i}\sum_{j\in i}\eta_{i}^{*}e^{-i\mathcal{A}_{ij}}\eta_j + \cdots \\
  &=& -t'\sum_{i}\sum_{j\in i}\Biggl\lbrack \eta_{i}^{*}\eta_{j}^{\phantom{*}} + \frac{\sin(A_{ij})}{A_{ij}}J_{ij}^{k}A_{ij}^{k} \nonumber \\
  &&  +\frac{1-\cos(A_{ij})}{A_{ij}^{2}}A_{ij}^{k}A_{ij}^{l}K_{ij}^{kl}\Biggr\rbrack + \cdots \nonumber
\end{eqnarray}
where $\mathcal{A}_{ij}=A_{ij}^kS^k$ is of the same kind as (\ref{GaugeField}) but constructed with the $S=1$ spin operators $S^k$ instead of the $S=\frac{1}{2}$ SU(2) generators $\frac{1}{2}\sigma^k$, and $A_{ij}^{\phantom{k}}=\sqrt{A_{ij}^kA_{ij}^k}$ (the summation over repeated upper indices is implied). The compact form of this coupling on the lattice has an additional term in the $S=1$ representation involving bond scalars:
\begin{equation}\label{BondScalars}
K_{ij}^{kl} = -\frac{1}{4}\Bigl(\eta_{i}^{*}\lbrace S^{k},S^{l}\rbrace\eta_{j}^{\phantom{*}}
                               +\eta_{j}^{*}\lbrace S^{k},S^{l}\rbrace\eta_{i}^{\phantom{*}}\Bigr) \ ,
\end{equation}
where $k,l\in\lbrace x,y \rbrace$. Note that $K_{ij}^{kl}=K_{ji}^{kl}=K_{ij}^{lk}$ are symmetric in both lower and upper indices. Only $K^{xx}$ and $K^{yy}$ couple to the gauge field (\ref{GaugeField}), but this is enough to generate a lattice pattern shown in Fig.\ref{vl-op}, which breaks lattice translation and rotation symmetries. This is ultimately responsible for the spin-orbit energy gain and the stability of the $J^z$ vortex lattice. Quantum fluctuations inevitably lower the average density of singularities by vortex-antivortex annihilations and make room for vortex motion on the lattice, but cannot remove all singularities because that would produce a higher-energy pair-density-wave condensate.

\begin{figure}[!]
\subfigure[{}]{\includegraphics[height=1.3in]{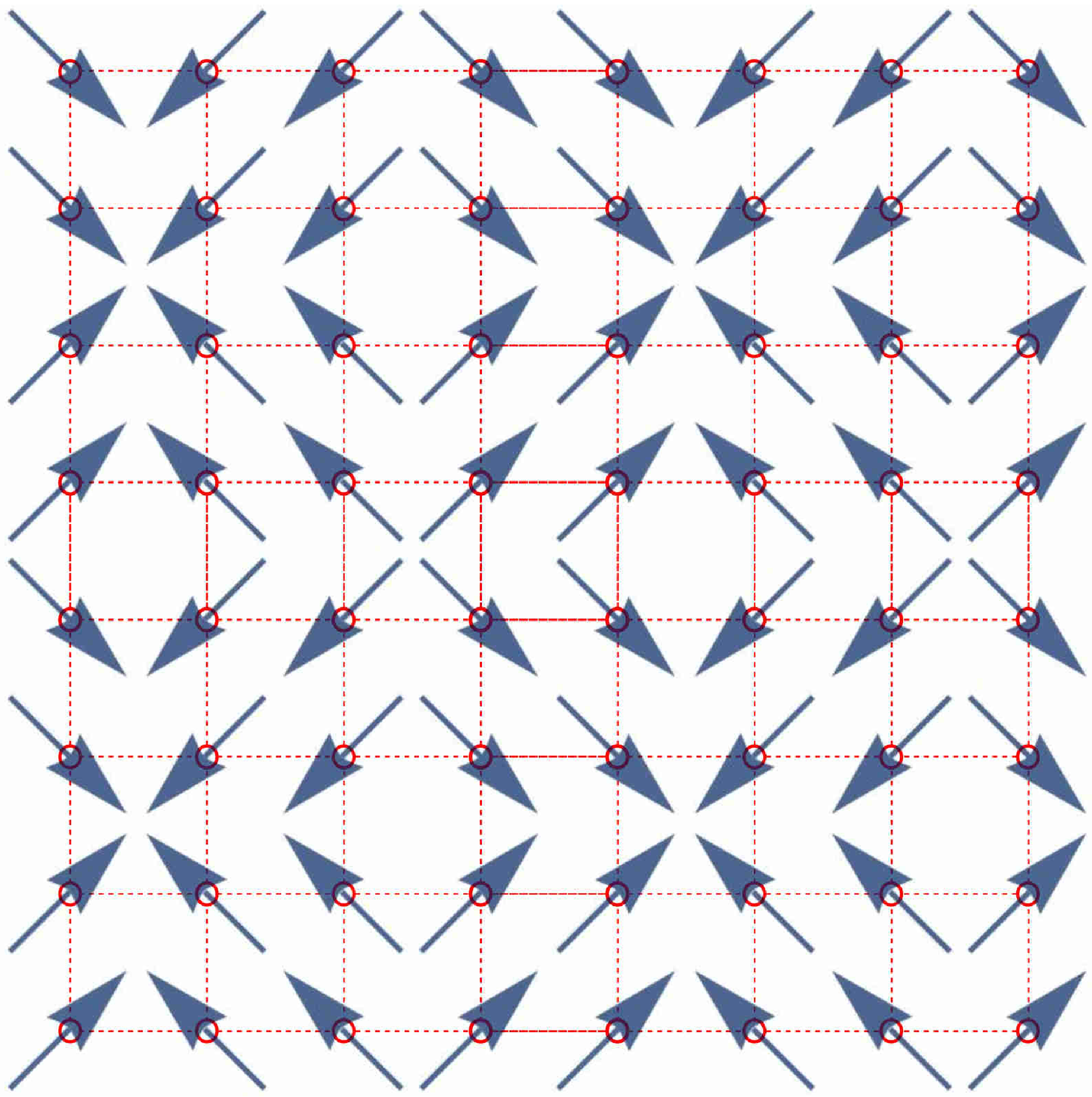}} \hskip 0.08in
\subfigure[{}]{\includegraphics[height=1.7in]{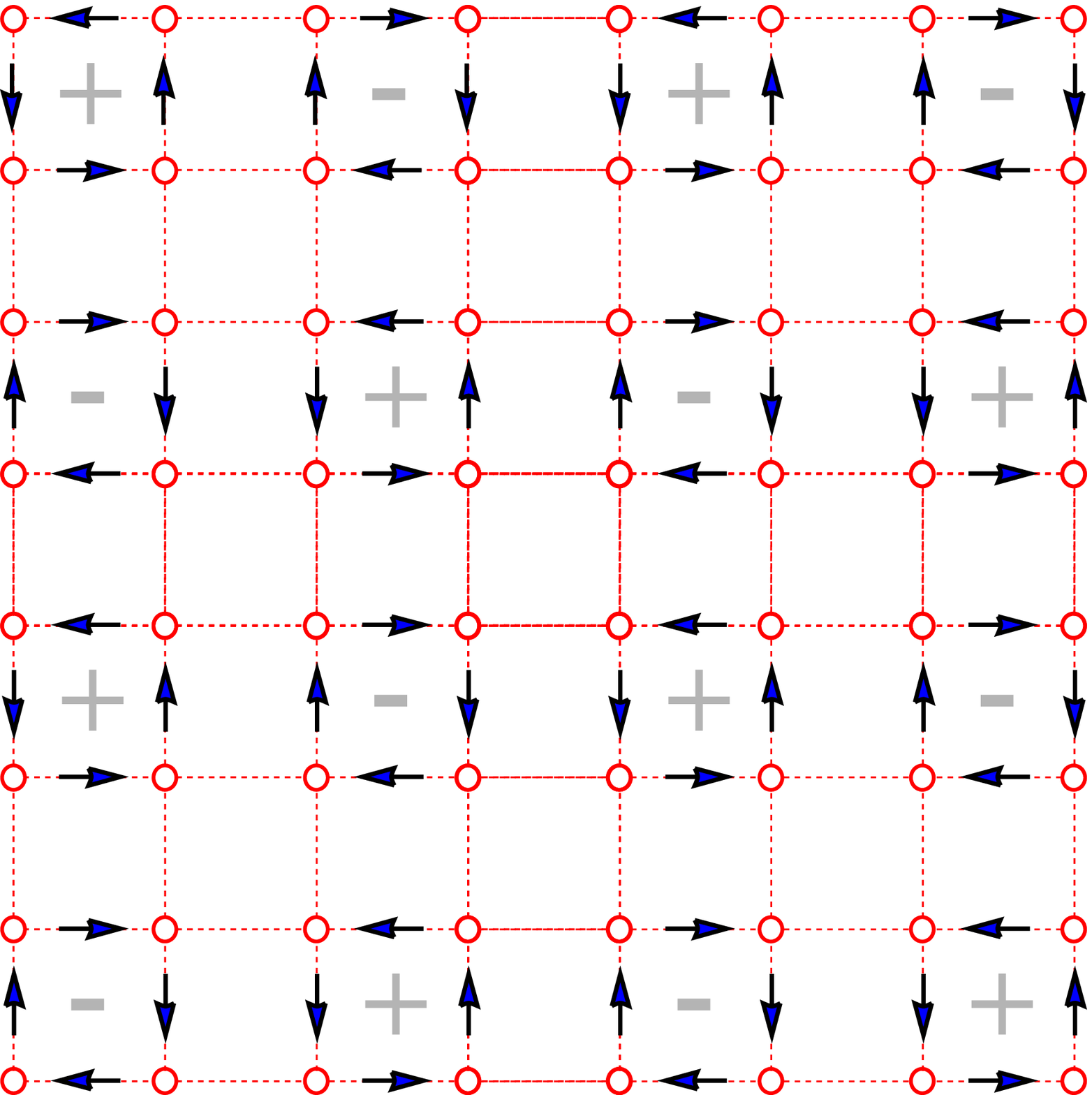}}
\caption{\label{vl2-op}The metastable vortex lattice with a $4\times 4$ unit-cell. (a) The phase $\theta$ of the order parameter component $\eta_\uparrow = |\bar{\eta}| e^{i\theta}$, and (b) $J^z$ spin-current. The plot details are the same as in Fig.\ref{vl-op}.}
\end{figure}

\begin{figure}[!]
\subfigure[{}]{\includegraphics[height=2.5in]{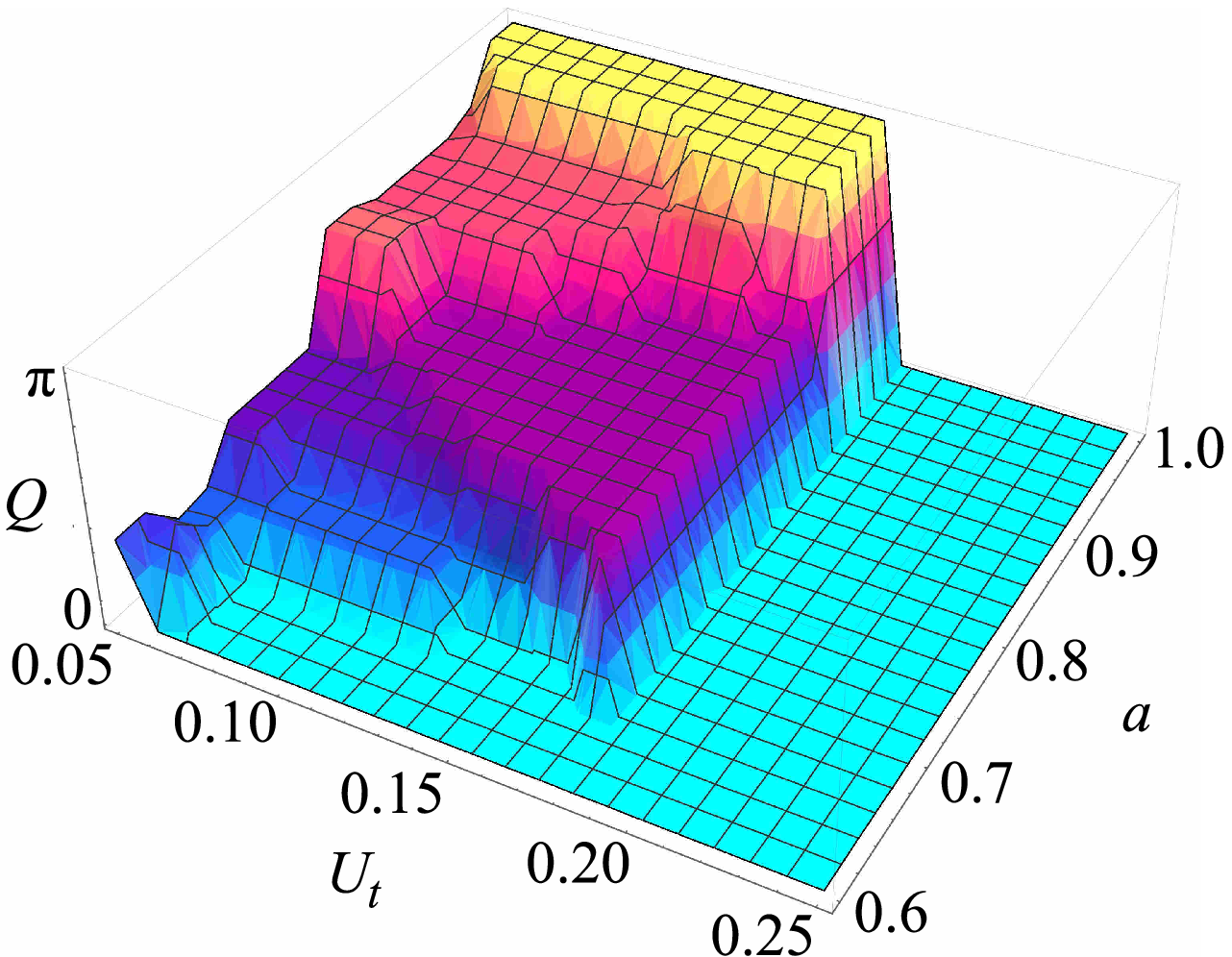}} \vskip 0.15in
\subfigure[{}]{\includegraphics[height=2.15in]{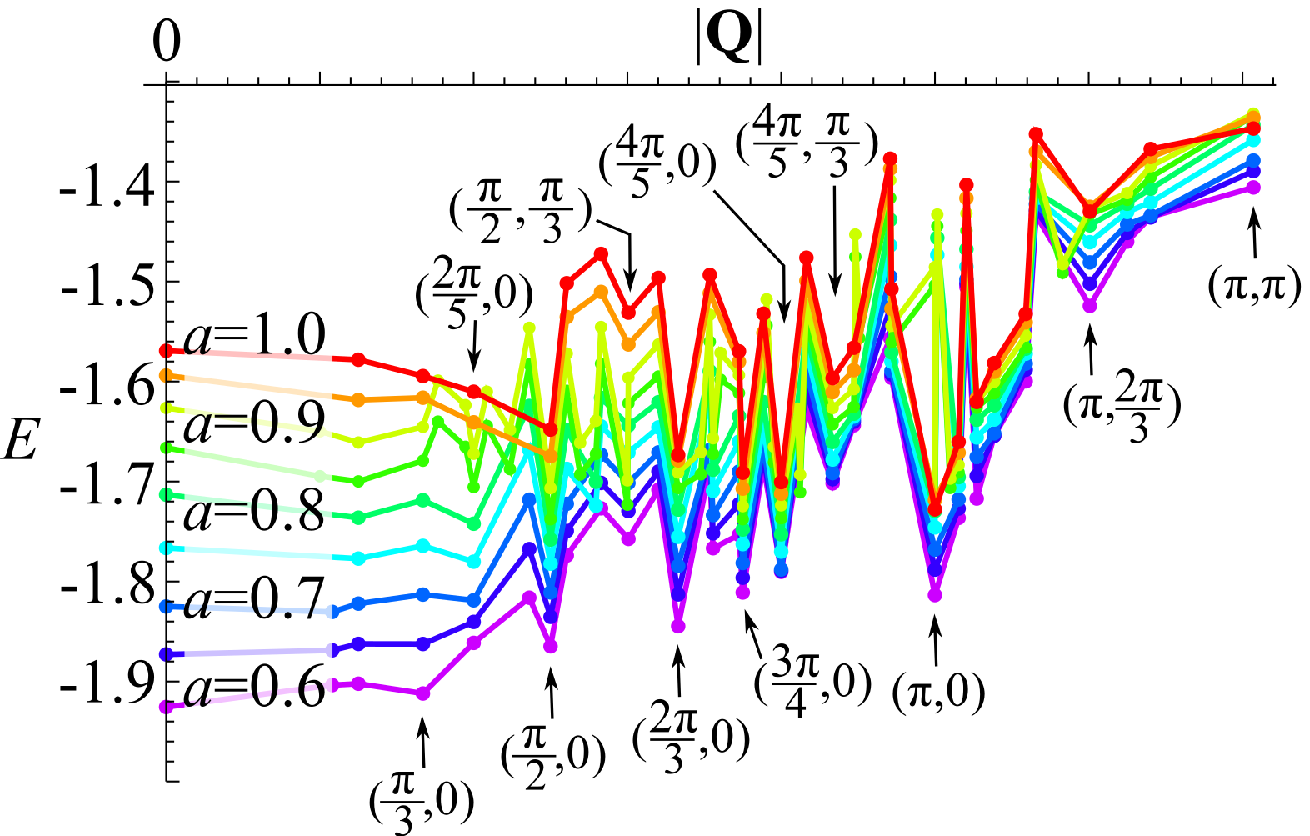}}
\caption{\label{Qvect}(a) The ordering wavevector ${\bf Q}=(Q,0)$ of the lowest energy PDW found by category-2 calculations ($U_{t0}=0.2$; $U_t$, $U_{t0}$ are expressed in the units of $t^{-1}$). (b) Energy (in the units of $t$) of stable and metastable PDW states as a function of the ordering wavevector ${\bf Q}$, parametrized by the spin-orbit coupling $a$ at $U_t=0.1$, $U_{t0}=0.2$.}
\end{figure}

The main competitors to the TR-invariant vortex lattices are various pair-density waves (PDWs). All found PDW states have ordering wavevectors along a lattice direction, e.g. ${\bf Q} = (Q,0)$. The magnitude $Q$ tends to grow with the spin-orbit coupling $a$ as expected. However, the lattice commensuration effects introduce large oscillations of the PDW energy $E_a(Q)$ as a function of $Q$ at sufficiently large values of $a$, as shown in Fig.\ref{Qvect}. The largest PDW energy dips are often found at $Q=2\pi/n$ for $n=1,2,3,4,5,6\dots$ near favorable values of $a$. The continuum limit is approached when $a$ is small and the ordering wavevector of the lowest-energy PDW is sufficiently small to be in the smooth (non-oscillatory) region of $E_a(Q)$. Among all these PDW states, the only one that ever wins as a mean-field ground state is the TR-invariant order at $Q=\pi/2$, seen for $0.7 \le a \le 0.875$. The next strongest PDW is also TR-invariant at large $a$ and $Q=\pi$, but it never wins. All other prominent PDWs found by category-2 calculations break the TR symmetry. Among the category-3 results, only the states at $Q\to 0$ (small $a$) and $Q\approx \pi$ (large $a$) can minimize energy, but not enough to win as ground states; still, only such PWS can be TR-invariant, so the states with TR symmetry consistently have lower energy than those that break it.

A separate category-1 calculation discovered additional metastable TR-invariant vortex lattices with a $4\times 4$ unit-cell, in the range $0.725\le a\le0.775$ where the ${\bf Q} = (\pi/2,0)$ PDW states are stable. The most prominent vortex pattern is the same checkerboard as before, but with a twice larger spatial separation between vortices shown in Fig.\ref{vl2-op}. It is interesting to note that the category-1 numerics fairly consistently found this particular type of local energy minimum instead of the global PDW minimum. Note that the explored phase-space was very large, since $16\times3\times2=96$ real quantities were needed to represent the order parameter on a $4\times 4$ unit-cell. In contrast, only $12$ real numbers are needed to specify the PDW order parameter of the true ground state discovered by the category-2 calculation. This probably indicates that vortex lattices have larger entropy (shallower and broader energy minimum) than the true PDW ground state. Quantum fluctuations beyond the mean-field approximation are expected to energetically enhance the states with larger entropy.

In the remaining scrutinized region $0.6 \le a \le 0.675$, the lowest energy condensates are found to be uniform at all $U_t, U_{t0}$. Even category-2,3 calculations settled for uniform states when $a \le 0.65$, despite having freedom to access smaller and incommensurate wavevectors. This is a consequence of being in the strong-coupling limit, with interaction energy scales $U_t^{-1}, U_{t0}^{-1}$ exceeding the bandgap $\Delta$. It was necessary to work in this limit in order to numerically resolve the tiny energy differences between various competing states (see Appendix), but the price to pay was a strong suppression of either $\eta_0$ or $\eta_{\uparrow\downarrow}$ order parameter components in any given phase (interactions compete for electrons to form pairs so the dominant interaction decides the pairing channel, in agreement with the first order transition at $U_t \approx U_{t0}$). Consequently, energy gain from the spin-orbit coupling requires finite bond scalars $K^{kl}$ (\ref{BondScalars}) when $\eta_0=0$. Orders of $K^{kl}$ are hard to arrange at larger wavevectors when $a$ becomes smaller, and their spin-orbit energy scales as $a^2$. Seeing finite-momentum ordering at smaller $a$ is probably possible only in the weaker coupling limit, which is not accessible with the current numerics. If interactions are not too strong, then the spin-orbit energy gain proportional to $a$ in (\ref{EffTheory}) can be sufficient to overcome the tendency of interactions to suppress $\eta_0$. The resulting states can have finite helical spin-currents $J^{x,y}$, and perhaps establish the vortex arrays envisioned in the Landau-Ginzburg analysis.

\subsection{Fluctuation effects}\label{secC}

The system studied here is two-dimensional, so one can anticipate significant corrections to the mean-field phase diagram. The continuous U(1) symmetry associated with superfluidity can be broken only at $T=0$, while it gives way to algebraic correlations at low $T>0$. All other symmetries of the model are discrete, and hence remain broken at low $T>0$.

At special values $a = n\frac{\pi}{2}$ of the spin-orbit coupling, the model (\ref{Model}) has another continuous symmetry. The spin-orbit coupling is absent if $n$ is even, so the corresponding ground-states are trivially uniform. However, the condensates at odd $n$ break the lattice translation symmetry, generally with a $2\times 2$ unit-cell. The $2\times 2$ vortex state that wins at $a<\frac{\pi}{2}$ becomes degenerate with the ${\bf Q} = (\pi,0)$ PWS and infinitely many other states when $a=\frac{\pi}{2}$. This follows from the symmetry (\ref{Sym2}), which allows free global rotations of the staggered $\theta$ pattern from Fig.\ref{vl-op}(b) in the opposite directions $\theta_A=-\theta_B$ on the sublattices A and B. Since (\ref{Sym2}) is a continuous symmetry at $a=\frac{\pi}{2}$, the degenerate condensates (e.g. vortex lattice and PWS) can be long-range ordered only at $T=0$ and become algebraically correlated at low finite temperatures. Note that this special symmetry is unrealistic because it requires the absence of next-nearest and further-neighbor hopping in the model.

The mean-field energy advantage of the vortex lattice over its related PWS gradually disappears as $a \to \frac{\pi}{2}$. Since the related PWS state is obtained by annihilating all singularities of the vortex lattice, one may expect prominent local quantum fluctuations that annihilate and re-create vortex-antivortex pairs when $a$ is close to $\frac{\pi}{2}$. The annihilations reduce the average density of vortices, and hence create room for vortex motion on the lattice. One possible outcome is the condensation of vortices, which by duality leads to a Mott insulator \cite{Fisher1989, Fisher1989a}. However, if vortices fail to condense due to their large density, still comparable to the density of particles, then a melted vortex lattice is an incompressible quantum liquid with topological order -- analogous to fractional quantum Hall states, but without TR symmetry breaking and quantized spin-Hall response \cite{Nikolic2012}.

The crystal lattice helps to pin the vortices, so a quantum phase transition between the vortex array condensate and a band-insulator can be of 2nd order. The \emph{mean-field} phase diagram features precisely such a transition when the intrinsic fermion bandgap $\Delta$ becomes sufficiently large (comparable to the pairing energy scale). If vortices must rely only on mutual interactions to establish a long-range positional order, then a quantum Lindemann criterion \cite{nikolic:144507} suggests that the mentioned 2nd order transition must be preempted by a 1st order quantum melting of the vortex array. Such melting occurs because the kinetic energy cost of localizing vortices exceeds the potential energy cost of deforming the vortex array when the order parameter amplitude becomes sufficiently small. For this reason, we expect that the actual phase diagram contains an intermediate vortex liquid phase between a vortex lattice condensate and a band-insulator.

Obviously, more analysis going beyond the mean-field approximation is needed to establish the accurate phase diagram of this two-dimensional system. However, the mean-field approximation is sufficient to reveal the existence of vortex ground states in the phase diagram of certain microscopic models like (\ref{Model}), which ultimately could be either crystalline or liquid. It is also adequate for establishing or confirming the predicted structure of vortices on the lattice, i.e. the patterns of charge and spin currents. The latter is a very useful information for anticipating the possible kinds of fractional quasiparticle statistics in putative incompressible vortex liquids. Field-theoretical arguments generally point to a particle-flux attachment in incompressible quantum liquids \cite{Nikolic2012}, so the non-trivial exchange statistics of quasiparticle excitations follows from the Aharonov-Bohm phase that one particle acquires on its path around the flux quanta attached to another quasiparticle. This simple principle may need to be combined with some deeper understanding of vortex dynamics in order to classify the possible topological orders in this system.

\section{Conclusions}\label{secConcl}

We studied the mean-field phase diagram of a simple two-dimensional model of attractively interacting fermions on the square lattice. The fermions experience a strong Rashba spin-orbit coupling in the manner characteristic for two-dimensional topological insulators. We discovered that a TR-invariant dense array of vortices has the lowest energy at the mean-field level among many competing orders in a large section of the phase diagram. This state is an analogue of the Abrikosov vortex lattice for the particular type of the SU(2) Yang-Mills flux that realizes the Rashba spin-orbit coupling. Other competing orders include pair-density waves and plane-wave states, some of which break the TR-symmetry. The energy competition between these states is fierce, so we expect that quantum and thermal fluctuations can easily destabilize some orders in favor of incompressible quantum liquids.

It should be pointed out that much of the observed phase diagram features are not universal. The phase transitions between the found competing orders are 1st order, and hence model-dependent. A somewhat different model may suppress or enhance the vortex states. Sorting out which microscopic details are favorable to vortex states is not a goal of the present work, and is probably not worth pursuing at present when correlated two-dimensional TIs are not experimentally available. However, our model is the simplest description of the physics that can be realistically expected in interacting TI quantum wells, on the boundaries of Kondo TIs, or engineered with cold atoms. In that sense, TI quantum wells are a promising system that may be able to realize the vortex arrays or liquids that we discussed in this paper.

{\bf Acknowledgements.} This work was supported by the National Science Foundation under Grant No. PHY-1205571, and also in part by the National Science Foundation under Grant No. PHY-1066293 with hospitality of the Aspen Center for Physics.

\appendix

\section{The competition between singlet and triplet pairing}

A more realistic model needs to include interactions that can lead to pairing in singlet channels. To that end, we can add the following interaction to the Hamiltonian (\ref{Hint}):
\begin{eqnarray}\label{Hint2}
H_{\textrm{int}}' \!\!&=&\!\! \sum_{i}\Biggl\lbrack
    U_{s+}|\phi_{i+}|^2+U_{s-}|\phi_{i-}|^2+U_{s0}|\phi_{i0}|^2 \nonumber \\
&&  +\phi_{i+}^{*}\psi_{i+\uparrow}^{\phantom{*}}\psi_{i+\downarrow}^{\phantom{*}}
    +\phi_{i-}^{*}\psi_{i-\uparrow}^{\phantom{*}}\psi_{i-\downarrow}^{\phantom{*}} \\
&&  +\phi_{i0}^{*}\frac{\psi_{i+\uparrow}^{\phantom{*}}\psi_{i-\downarrow}^{\phantom{*}}
        -\psi_{i+\downarrow}^{\phantom{*}}\psi_{i-\uparrow}^{\phantom{*}}}{\sqrt{2}}+h.c.\Biggr\rbrack \ . \nonumber
\end{eqnarray}
The order parameters $\phi_{i+}$ and $\phi_{i-}$ are singlet pairs of two electrons on the same Dirac cone, $\tau=+1$ or $\tau=-1$, while $\phi_{i0}$ is the inter-cone singlet. In general, no special symmetry governs the values of the interaction couplings $U_{s+}, U_{s-}, U_{s0}$.

\begin{figure}
\includegraphics[height=2.1in]{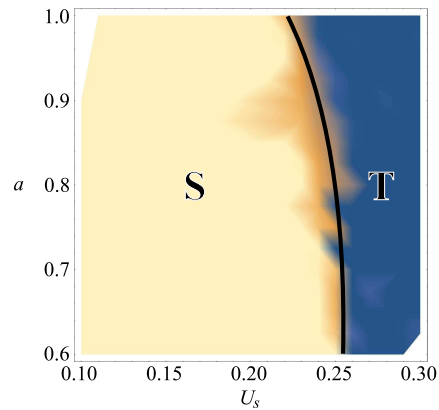}
\caption{\label{singlet}A typical first-order phase boundary between singlet (S) and triplet (T) orders, which illustrates that the triplet phases expands further with larger spin-orbit couplings $a$. The singlet phases are uniform, while the triplet phases exhibit a variety of non-uniform orders that were discussed before. The transitions that separate different singlet and triplet phases are not shown. In this plot, $U_{s+}=U_{s-}=U_s$ and $U_{s0}=0.2$, $U_t=0.17$, $U_{t0}=0.2$, all in units of $t^{-1}$.}
\end{figure}

Category-1 calculations were carried out to examine the competition between the singlet $\phi$ and triplet $\eta$ order parameters in the present model. A general conclusion is as expected: the spin channel with strongest coupling (smallest order parameter suppression $U$) wins. The transition between singlet and triplet orders is first order. The only interesting new feature of this model is that the spin-orbit coupling tends to strengthen and enhance the triplet orders. This is illustrated in the typical singlet-triplet phase diagram, Fig.\ref{singlet}.

Note that even the $\eta_0$-dominated phase in the region $U_t \gtrsim U_{t0}$ is enhanced by the spin-orbit coupling, despite being uniform. This is a consequence of the quadratic spin-orbit effect in (\ref{EffTheory}): both $K^{xx}$ and $K^{yy}$ are finite. One consequence is that the interesting vortex lattice and PDW phases are not helped by the spin-orbit coupling in their competition with the uniform $\eta_0$ phase. However, all triplet orders have a certain advantage over the singlet orders when the spin-orbit coupling is large.

\section{The competition between pair density waves}

\begin{figure*}
\includegraphics[height=2.2in]{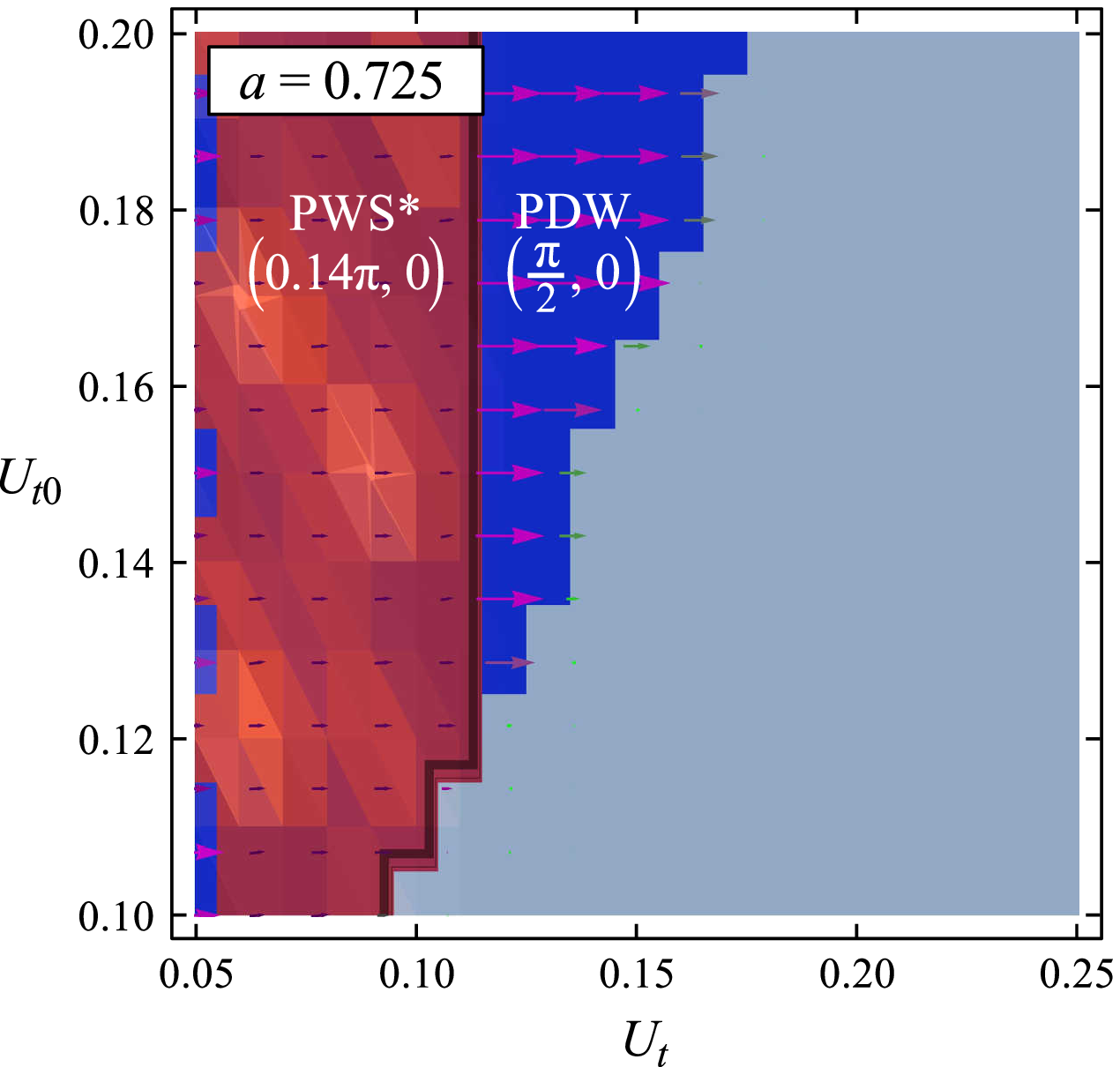} \hskip 0.03in
\includegraphics[height=2.2in]{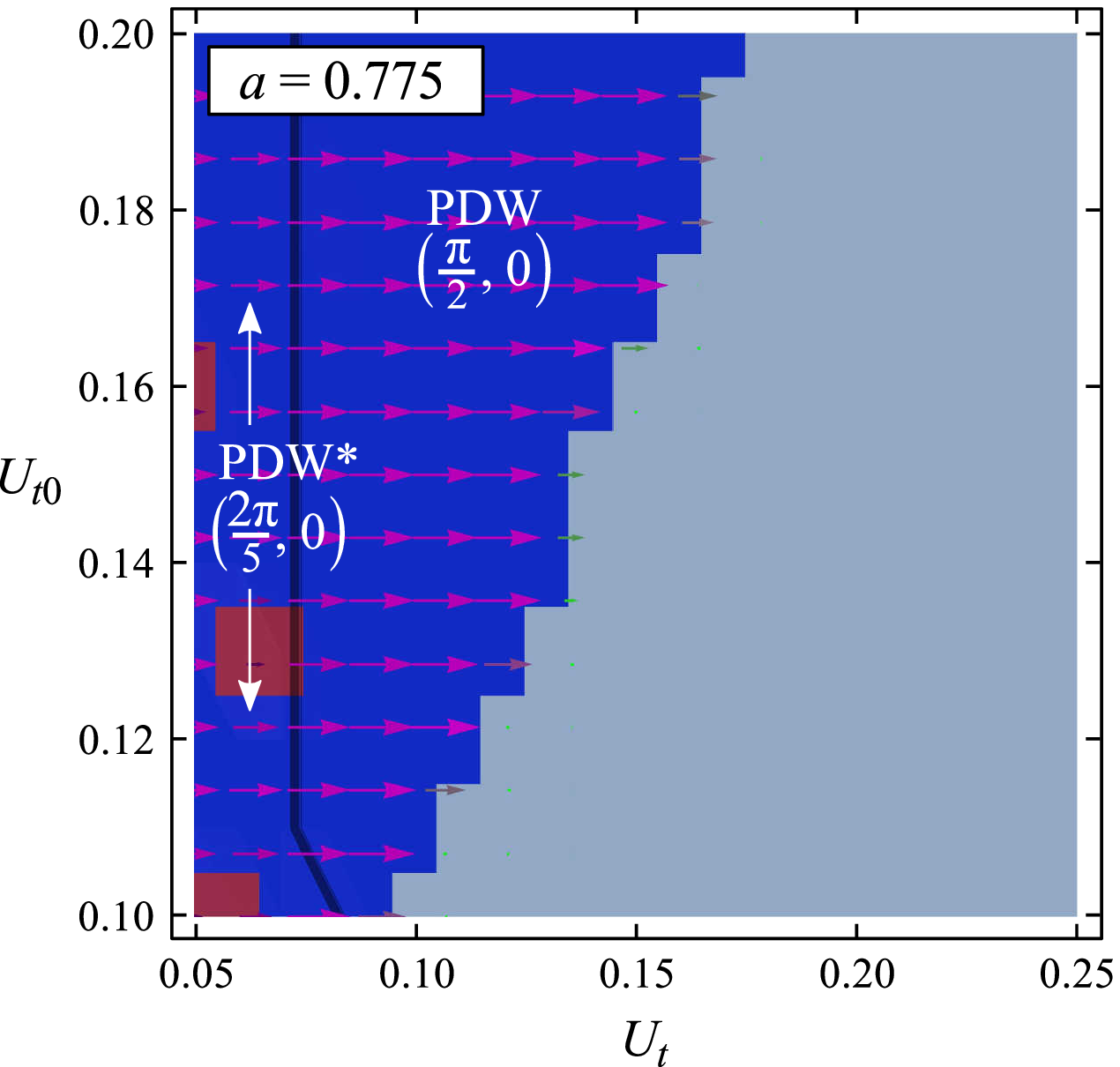} \hskip 0.03in
\includegraphics[height=2.2in]{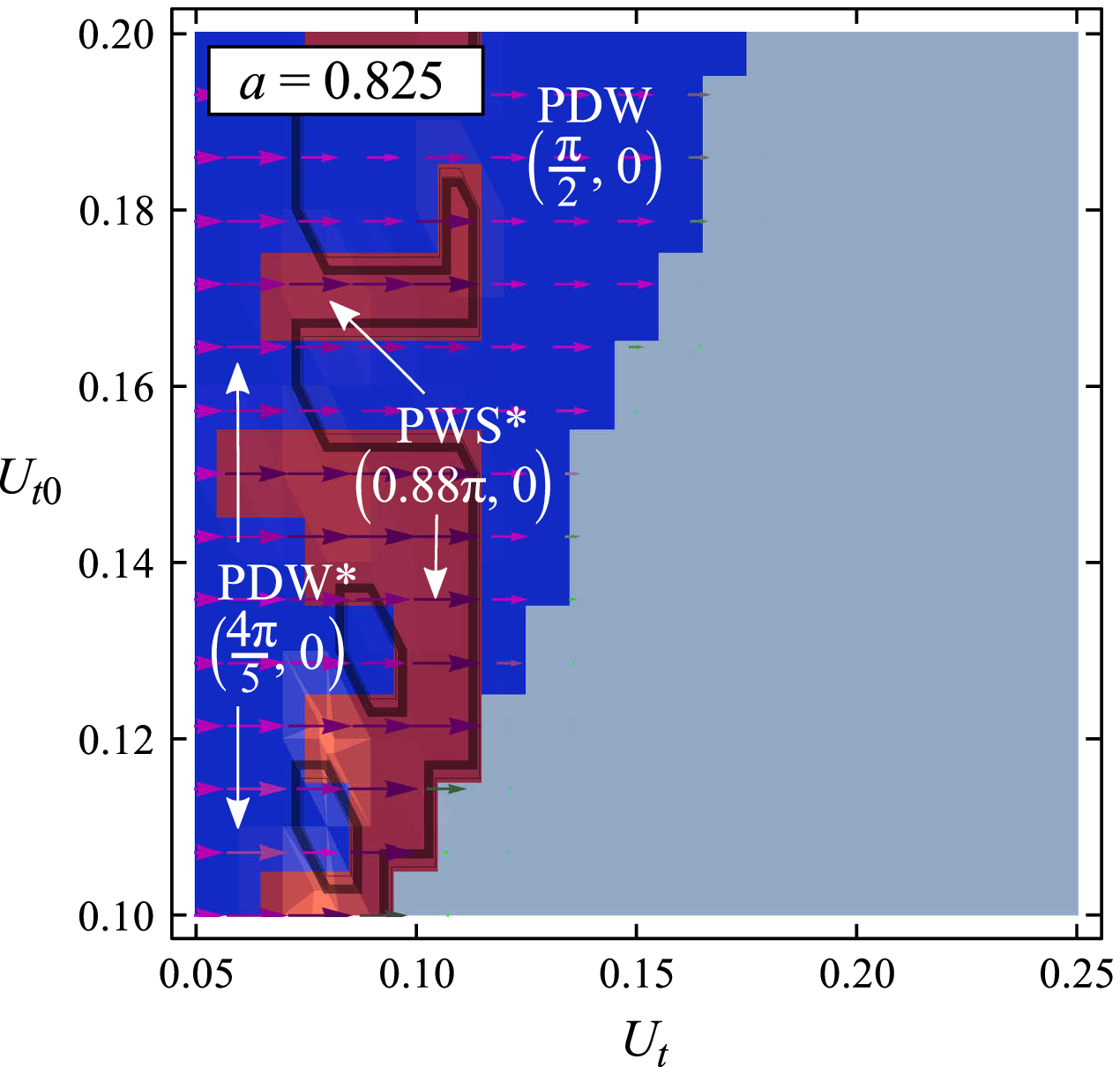}
\includegraphics[height=2.2in]{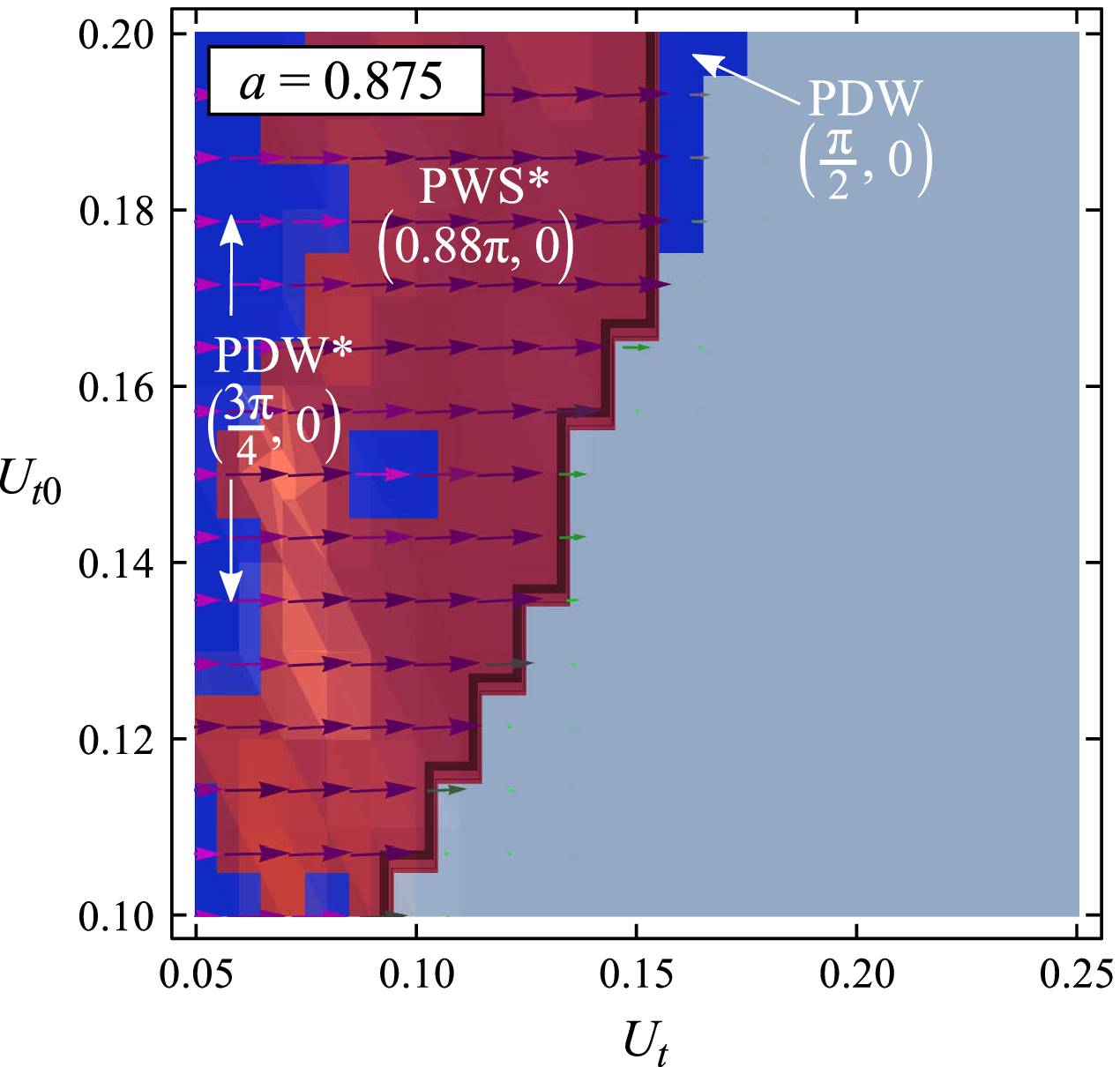} \hskip 0.03in
\includegraphics[height=2.2in]{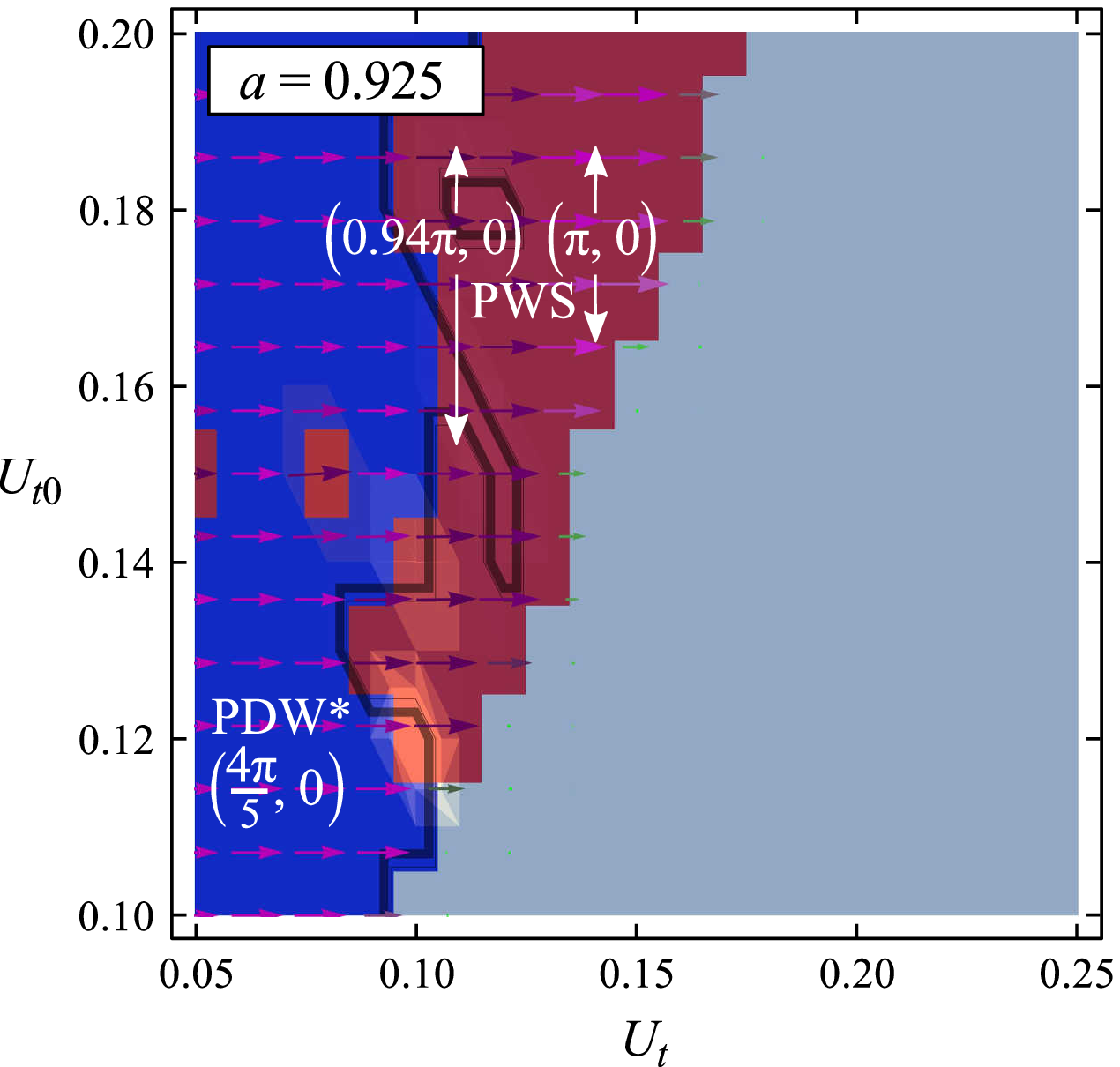} \hskip 0.03in
\includegraphics[height=2.2in]{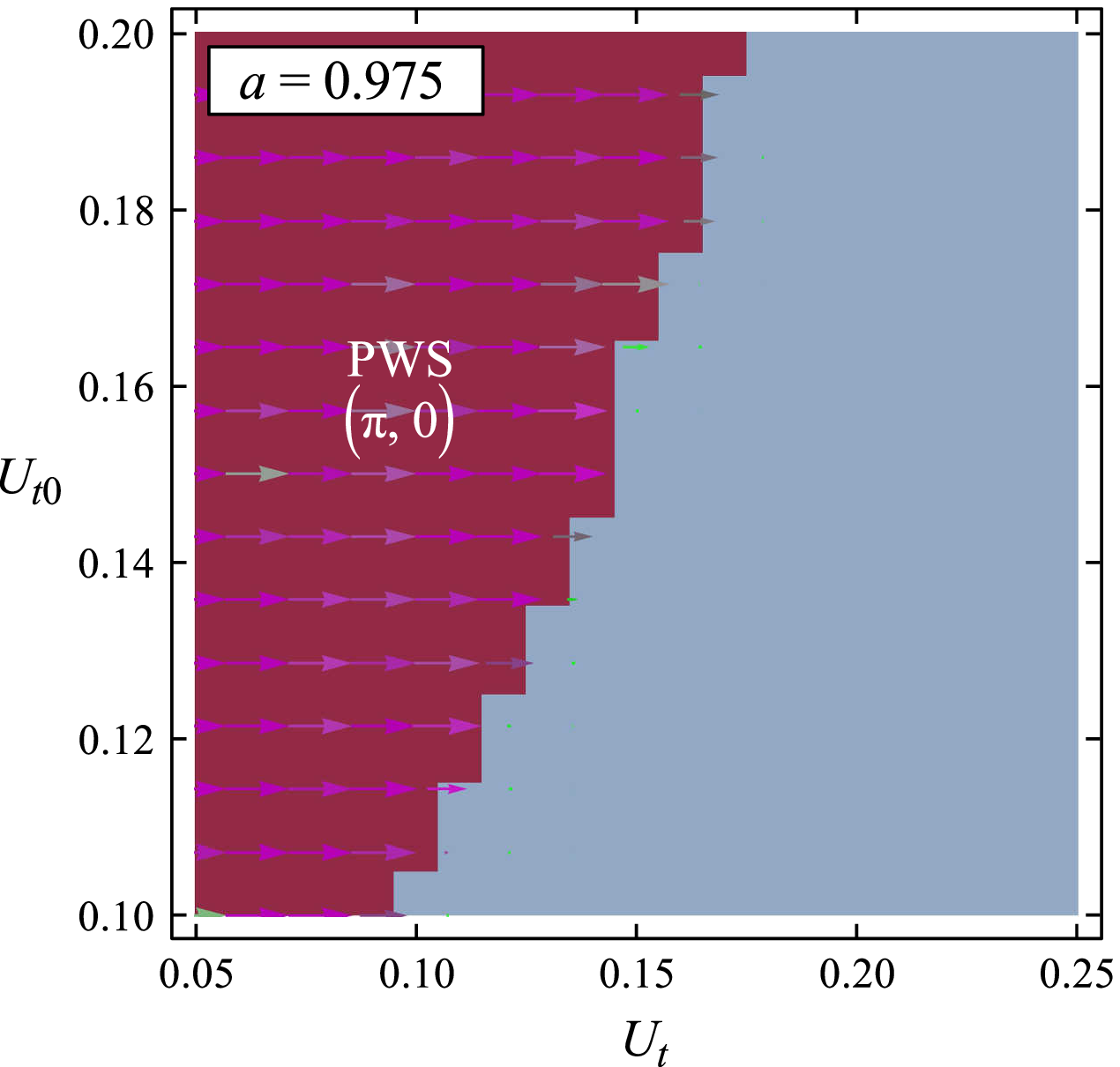}
\caption{\label{phdiag2}A sample of the mean-field phase diagram of metastable pair density waves (PDW) and plain wave states (PWS), obtained by comparing the best results of category-2 and category-3 calculations. Stars indicate TR symmetry breaking, and thick black lines indicate a transition between TR-invariant and TR-broken states. Some regions of the phase diagram are noisy due to numerical minimization errors in a frustrated energy landscape, especially when the same PWS order at ${\bf Q}=(\pi,0)$ is detected by both category 2 and 3 runs. However, almost all noise issues are resolved by the winning vortex lattice phase, not shown here (see Fig.\ref{phdiag1}).}
\end{figure*}

Fig.\ref{phdiag2} shows the phase diagram of competing PDW and PWS states. Arrows indicate (by length and orientation) the ordering wavevector ${\bf Q}$ of PWS and PDW states that were found to have the lowest energy by a category-2 or -3 calculation. Some PDW/PWS states were discovered by a category-1 calculation, in which case they are not accompanied by arrows; being defined in a $2\times 2$ unit-cell, their wavevector is then ${\bf Q} \in \lbrace (0,\pi), (\pi,0), (\pi,\pi) \rbrace$. The arrow color (not clearly visible everywhere even in the online color plots) indicates the relative amplitudes of the triplet spinor components. The amounts of R(red), G(green), B(blue) pigments in the RGB arrow color is set by the normalized amplitudes of $\eta_{\uparrow}$, $\eta_{0}$, and $\eta_{\downarrow}$ components respectively. Therefore, purple=red+blue arrows (hard to see) denote the PDW/PWS states dominated by $\eta_{\uparrow}$ and $\eta_{\downarrow}$ in roughly equal amounts when $U_t \lesssim U_{t0}$.

The sampled points in the phase diagram are represented by painted blocks with the following color scheme. Green colors are vortex arrays with four singularities in the unit-cell of $2\times2$ lattice sites. Blue colors are PDW states with non-zero order parameter amplitudes at two physically different commensurate wavevectors. Hence, a PDW state breaks translation symmetry. If a blue block is accompanied by an arrow, then the two wavevectors are ${\bf Q}$ and $-{\bf Q}$ (and not high-symmetry ones in the 1st Brillouin zone). Red colors are PWS states with a non-zero amplitude at a single wavevector ${\bf Q}$ (which can be a non-zero high-symmetry wavevector). A PWS respects translation symmetry up to a global U(1) gauge transformation. Light purple represents a conventional uniform superfluid condensed at ${\bf Q}=0$, and white is a band-insulator with a vanishing order parameter. Regions with spontaneous magnetization are painted with a lighter color, but follow the same color scheme. All states with spontaneous TR symmetry breaking are separated by a pair of black lines, where the thinner line faces the TR-invariant state. The size of blocks reveals the resolution at which the phase diagram was scanned.

The dense vortex array (Fig.\ref{vl-op}) discovered by the category-1 calculation can be regarded as a special kind of PDW states. Its lattice Fourier transform has finite amplitudes only at the wavevectors $(\pi,0)$ and $(0,\pi)$. The metastable vortex array in a $4\times 4$ unit-cell (Fig.\ref{vl2-op}) has equal finite Fourier amplitudes at the wavevectors $(\pm\frac{\pi}{2},0)$ and $(0,\pm\frac{\pi}{2})$. The qualitative distinction from the conventional PDW states is made primarily by the presence of circulating currents. The vorticity of such currents is counted by first extrapolating the order parameter to real-valued coordinates, and then measuring the winding number of its resulting smoothly varying phase on the continuous paths around lattice plaquettes. This particular interpretation of vorticity in the \emph{lattice} order parameter (whose phase may jump from one site to another) is somewhat arbitrary, but the pattern of circulating currents is physically measurable and imprints an Aharonov-Bohm phase on any degree of freedom that couples to it.

\section{Numerical procedure and error estimates}

The three category calculations carry out the same numerical procedure, and share the same code for common tasks. They mainly differ by the manner they define and handle the order parameter. Each type of the order parameter requires its own representation of the Bogoliubov de-Gennes (BdG) Hamiltonian for quasiparticle excitations.

The category-1 search specifies the order parameter on a real-space unit-cell of $l_x \times l_y$ sites. Both $l_x$ and $l_y$ are fixed before the search. The BdG Hamiltonian is represented in the Bloch-state basis corresponding to this unit-cell, while the unit-cell internals are handled in real-space. The category-2 search specifies its order parameter by Fourier triplet amplitudes at the commensurate wavevectors ${\bf Q}$ and ${-\bf Q}$ (which may be physically the same if ${\bf Q}$ is a high-symmetry wavevector). Its BdG Hamiltonian is also represented in momentum space, using the largest reduced 1st Brillouin zone allowed by ${\bf Q}$. The category-3 BdG Hamiltonian is the simplest and structurally similar to that constructed for uniform condensates in the momentum space (pairing occurs between two fermions at momenta displaced by ${\bf Q}$). The technical details of these BdG Hamiltonians are presented in the next section. 

\begin{figure}
\includegraphics[width=3.2in]{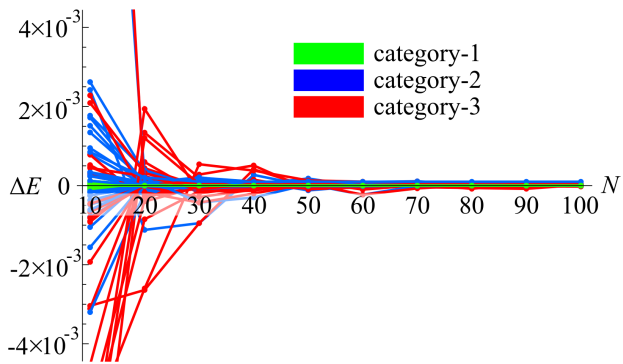}
\caption{\label{errors}The difference $\Delta E_i = E_i - E$ between energy $E_i$ calculated by the category $i=1,2,3$ method and the reference accurate energy $E$ of 36 randomly chosen states, plotted as a function of the number $N$ of sampled points per spatial dimension for the 1st Brillouin zone integration (see text for the explanation). Each connected sequence of circles in a single color corresponds to a different order parameter (with varied ordering wavevectors and unit-cell sizes), but the energy of every chosen state was calculated with all category methods that can handle such a state. The category-1 energy at $N=100$ is used as a reference $E$. The plot is meant to illustrate the pace and manner in which the calculated energies converge to the accurate values.}
\end{figure}

Once a BdG Hamiltonian is constructed, all three category searches diagonalize it and then obtain the energy of the many-body system (formally non-interacting after the mean-field approximation) by integrating out the single-particle BdG spectrum up to the Fermi energy $\mu$ in the appropriately reduced 1st Brillouin zone. Since the model has well-defined cut-offs, this numerical integration is the only source of errors in the calculations of energy. The integration employs Gauss-Legendre integration formulas, based on sampling the integrated functions at a set of discrete points in the 1st Brillouin zone; the sampled function is interpolated by a polynomial that can be integrated out exactly. One controls the absolute error of the calculated energy by the number $N$ of sampled points per spatial dimension in the 1st Brillouin zone (1BZ), as shown in Fig.\ref{errors}.

The ground state search employs the simplex method to minimize the state energy as a function of state parameters. Each category represents the superfluid state by its own limited set of $n$ variable parameters, as discussed earlier. The simplex method keeps track of the function values at $n+1$ points (vertices of the simplex) in the abstract $n$-dimensional variable space. One vertex of the simplex is moved to a new position in each iteration of the method, according to a set of rules that gradually make the simplex drift toward and shrink around a local minimum in the parameter space. A search for the local minimum begins by specifying an initial ``seed'' state (the center of the simplex) and the overall size of the simplex (the distance between vertices). The search ends after a number of iterations when the simplex shrinks below a requested minimum size. The typical accuracy of the obtained order parameters is $\sim 0.1 \%$ of the absolute order parameter magnitude. In order to speed up the phase diagram mapping, all discovered ground states are recorded for future use as ``seeds'' on the nearby points of the phase diagram. A few independent phase diagram scans, and many independent calculations on arbitrarily selected points, were conducted with random ``seeds'' in an attempt to reduce the danger of falling into metastable states.

The search for the minimum-energy states was carried out with a reasonable level of accuracy that did not cost an extreme computation time. This was $N_1=8$, $N_2=8$ and $N_3=100$ sampled 1BZ points per dimension for the categories 1, 2, 3 respectively. The category 1 is found to be very accurate even at $N\sim 10$, while the categories 2 and 3 reach a similar accuracy only at $N>50$ (see Fig.\ref{errors}). For the great majority of these order parameters, the three category calculations of energy agree to a fraction of a percent of the absolute energy value with $N=100$, making absolute errors of about $(1-2) \times 10^{-4} t$.
Energy minimization at high accuracy is prohibitively slow (except in category-3), so the optimal order parameters were obtained by using a manageable accuracy $N=8$. However, the energies of discovered order parameters were recalculated at the end with high accuracy $N=100$. All presented conclusions about the phase diagram follow from the comparisons of such accurate order-parameter energies in different categories. It is very unlikely that a highly accurate minimization would shift the order parameters enough to qualitatively change the phase diagram.

\begin{figure}[!]
\includegraphics[width=3in]{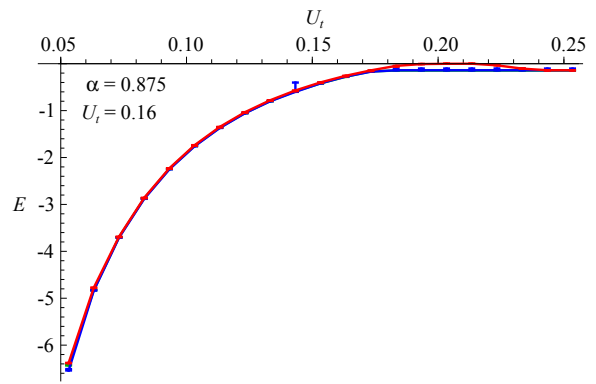}
\caption{\label{fe-abs}A typical energy scale $E$ of the lowest energy condensates in all three categories on a path through the phase diagram.  The units for $E$ are given by the hopping $t$. The best states in the category 1,2,3 are plotted by the green, blue and red lines respectively, but these lines can hardly be distinguished because they coincide when plotted on the overall energy scales.}
\end{figure}

The typical energy scale of the paired ground states is shown in Fig.\ref{fe-abs}. This plot compares the best condensates found in all three categories, but their energies cannot be distinguished on the absolute energy scale. This illustrates a significant frustration of the model and a fierce competition between different orders. The absolute energy difference between competing orders is often just $\sim 10^{-3} t$, but this is sufficiently larger than the absolute error to reliably  determine the ground state.

The category-2 search is particularly complicated and tedious because one variable property of the order parameter is its commensurate ordering wavevector ${\bf Q}$. The discrete components of ${\bf Q}$ must be varied outside of the simplex energy minimization. We systematically scan for the metastable states at different ${\bf Q}$ in the order that generally favors smaller unit-cells. The category-2 algorithm searches several times for an energy minimum at every checked ${\bf Q}$. These searches are at first done with a few restricted forms of the category-2 order parameter that were observed to often minimize energy. Then the best found state is used as a ``seed'' for the final least-restricted category-2 search. Each search starts with the same finite simplex size, to look for the minimum within a significant neighborhood of the ``seed''. Only the best overall result is ultimately reported for each ${\bf Q}$ as a metastable state energy. New wavevectors ${\bf Q}$ are checked down the line until the lowest discovered metastable energy has not changed in a while. This often reaches the unit-cells with up to $4\times 5$ or $3\times 7$ sites.

The category-1 search also tests several candidate states at each point in the phase diagram. It first obtains the lowest possible energy of all PDW/PWS states that can fit in the specified unit-cell of $l_x\times l_y$ sites (this step was skipped in the calculations on the $4\times 4$ unit-cell). Then it constructs restricted order parameters that loosely correspond to type-1 and type-2 states discussed in Ref. \cite{Nikolic2014}, and minimizes their energy. Each new attempt uses the previous best result as a ``seed''. The final minimization is restricted only by the specified unit-cell dimensions, and only the best overall state is ultimately reported.

Given this aggressive procedure, one can be reasonably confident that the ordered states with small unit-cells have been captured correctly. For example, we can be reasonably confident that a PDW or PWS with a $2\times2$ unit-cell is not the global energy minimum in the regions of the phase diagram where the TR-invariant $2\times2$ vortex lattice was reported as a consistent winner. We can be still fairly confident that no PDW/PWS state with a slightly larger unit-cell is the global minimum in these regions. Searching as long as it takes for the most optimal ordering wavevector ${\bf Q}$ of a PDW is futile, so we can never rule out the existence of global PDW minimums with larger unit-cells. However, considering also the local quantum or thermal fluctuations, the ordered states with larger unit-cells are almost invariably destabilized more than the states with smaller unit-cells. Fluctuations are certain to be prominent in our very frustrated system.

\section{Representations of the Bogoliubov de-Gennes Hamiltonian}

\subsubsection*{Category 1}

The category-1 periodic order parameter is specified by complex triplet amplitudes $\eta = (\eta_\uparrow, \eta_0, \eta_\downarrow)$ at every site of a fixed unit-cell. The fermion fields $\psi$ are organized into four-component spinors, but representing their triplet pairing requires an eight-component Nambu spinor at each site $i$:
\begin{equation}\label{Nambu1}
\Psi_{i}^{T}=\left(\begin{array}{cccccccc}
\psi_{i+\uparrow}^{\phantom{*}} & \psi_{i+\downarrow}^{*} & \psi_{i+\downarrow}^{\phantom{*}} & \psi_{i+\uparrow}^{*} & \psi_{i-\uparrow}^{\phantom{*}} & \psi_{i-\downarrow}^{*} & \psi_{i-\downarrow}^{\phantom{*}} & \psi_{i-\uparrow}^{*}\end{array}\right) \nonumber
\end{equation}
Expressing all terms of (\ref{H0}) and (\ref{Hint}) that contain fermion fields as a product of the Nambu spinors $\Psi^\dagger$ and $\Psi$ yields the BdG Hamiltonian:
\begin{equation}\label{HBdG1}
H_{\textrm{BdG}}^{\phantom{\dagger}} = \sum_{ij} \Psi_i^\dagger h_{ij}^{\phantom{\dagger}} \Psi_j^{\phantom{\dagger}} \ ,
\end{equation}
where the matrices $h_{ij}$ are defined only locally ($i=j$):
\begin{eqnarray}
h_{ii} = \left(\begin{array}{cccccccc}
-\mu & 0 & 0 & 0 & \Delta & -\frac{\eta_{0}^{\phantom{*}}}{\sqrt{2}} & 0 & -\eta_{\uparrow}^{\phantom{*}}\\
0 & \mu & 0 & 0 & \frac{\eta_{0}^{*}}{\sqrt{2}} & -\Delta & \eta_{\downarrow}^{*} & 0\\
0 & 0 & -\mu & 0 & 0 & -\eta_{\downarrow}^{\phantom{*}} & \Delta & -\frac{\eta_{0}^{\phantom{*}}}{\sqrt{2}}\\
0 & 0 & 0 & \mu & \eta_{\uparrow}^{*} & 0 & \frac{\eta_{0}^{*}}{\sqrt{2}} & -\Delta\\
\Delta & \frac{\eta_{0}^{\phantom{*}}}{\sqrt{2}} & 0 & \eta_{\uparrow}^{\phantom{*}} & -\mu & 0 & 0 & 0\\
-\frac{\eta_{0}^{*}}{\sqrt{2}} & -\Delta & -\eta_{\downarrow}^{*} & 0 & 0 & \mu & 0 & 0\\
0 & \eta_{\downarrow}^{\phantom{*}} & \Delta & \frac{\eta_{0}^{\phantom{*}}}{\sqrt{2}} & 0 & 0 & -\mu & 0\\
-\eta_{\uparrow}^{*} & 0 & -\frac{\eta_{0}^{*}}{\sqrt{2}} & -\Delta & 0 & 0 & 0 & \mu
\end{array}\right) \nonumber
\end{eqnarray}
and for the pairs of nearest-neighbor sites $h_{i,i+\hat{x}} = -t h_x$, $h_{i,i+\hat{y}} = -t h_y$:
\begin{eqnarray}
h_x = \left(\begin{array}{cccccccc}
C_a & 0 & -S_a & 0 & 0 & 0 & 0 & 0\\
0 & -C_a & 0 & -S_a & 0 & 0 & 0 & 0\\
S_a & 0 & C_a & 0 & 0 & 0 & 0 & 0\\
0 & S_a & 0 & -C_a & 0 & 0 & 0 & 0\\
0 & 0 & 0 & 0 & C_a & 0 & S_a & 0\\
0 & 0 & 0 & 0 & 0 & -C_a & 0 & S_a\\
0 & 0 & 0 & 0 & -S_a & 0 & C_a & 0\\
0 & 0 & 0 & 0 & 0 & -S_a & 0 & -C_a
\end{array}\right) \nonumber
\end{eqnarray}
\begin{eqnarray}
h_y = \left(\begin{array}{cccccccc}
C_a & 0 & iS_a & 0 & 0 & 0 & 0 & 0\\
0 & -C_a & 0 & iS_a & 0 & 0 & 0 & 0\\
iS_a & 0 & C_a & 0 & 0 & 0 & 0 & 0\\
0 & iS_a & 0 & -C_a & 0 & 0 & 0 & 0\\
0 & 0 & 0 & 0 & C_a & 0 & -iS_a & 0\\
0 & 0 & 0 & 0 & 0 & -C_a & 0 & -iS_a\\
0 & 0 & 0 & 0 & -iS_a & 0 & C_a & 0\\
0 & 0 & 0 & 0 & 0 & -iS_a & 0 & -C_a
\end{array}\right) \nonumber
\end{eqnarray}
with $C_a = \cos(a)$, $S_a = \sin(a)$. Including interactions in the uniform singlet pairing channels of equation \ref{Hint2} amounts to adding
\begin{eqnarray}
h_{ii}' = \left(\begin{array}{cccccccc}
0 & -\phi_+^{\phantom{*}} & 0 & 0 & 0 & -\frac{\phi_{0}^{\phantom{*}}}{\sqrt{2}} & 0 & 0\\
-\phi_+^* & 0 & 0 & 0 & -\frac{\phi_{0}^{*}}{\sqrt{2}} & 0 & 0 & 0\\
0 & 0 & 0 & \phi_+^{\phantom{*}} & 0 & 0 & 0 & \frac{\phi_{0}^{\phantom{*}}}{\sqrt{2}}\\
0 & 0 & \phi_+^* & 0 & 0 & 0 & \frac{\phi_{0}^{*}}{\sqrt{2}} & 0\\
0 & -\frac{\phi_{0}^{\phantom{*}}}{\sqrt{2}} & 0 & 0 & 0 & -\phi_-^{\phantom{*}} & 0 & 0\\
-\frac{\phi_{0}^{*}}{\sqrt{2}} & 0 & 0 & 0 & -\phi_-^* & 0 & 0 & 0\\
0 & 0 & 0 & \frac{\phi_{0}^{\phantom{*}}}{\sqrt{2}} & 0 & 0 & 0 & \phi_-^{\phantom{*}}\\
0 & 0 & \frac{\phi_{0}^{*}}{\sqrt{2}} & 0 & 0 & 0 & \phi_-^* & 0
\end{array}\right) \nonumber
\end{eqnarray}
to $h_{ii}$. Next, we carry out the Fourier transform of (\ref{HBdG1}) by introducing a crystal momentum ${\bf k}$ that lives in the reduced 1st Brillouin zone $k_x \in (-\pi/l_x, \pi/l_x)$, $k_y \in (-\pi/l_y, \pi/l_y)$ corresponding to the order parameter's unit-cell. We must formally join $l_x l_y$ Nambu spinors (\ref{Nambu1}) into a single super-spinor, whose Fourier transform $\bar{\Psi}_{\bf k}$ corresponds to a Bloch state. The Fourier-transformed Hamiltonian is:
\begin{equation}\label{HBdG2}
H_{\textrm{BdG}} = \int\limits _{\textrm{1BZ}}\frac{d^{2}k}{(2\pi)^{2}}\,\bar{\Psi}_{{\bf k}}^{\dagger}\left(\hat{h}+\sum_{\mu}\left(\hat{c}_{\mu}e^{ik_{\mu}}+h.c.\right)\right)\bar{\Psi}_{{\bf k}}^{\phantom{\dagger}}
\end{equation}
$\hat{h}$ is constructed as a block-matrix with blocks $h_{ii}$ (diagonal) and $h_{ij}$ (off-diagonal) for every two sites $ij$ in the same unit-cell. $c_\mu$, $\mu \in \lbrace x,y \rbrace$ are the ${\bf k}$-dependent block-matrices made from $h_{ij}$ that connect the sites belonging to different unit-cells.

The full spectrum of the BdG Hamiltonian can now be obtained by diagonalizing the $8l_xl_y\times8l_xl_y$ matrix for any ${\bf k}$ in (\ref{HBdG2}). Since the Nambu representation had to be doubled for triplet pairing, the spectrum has an unphysical double degeneracy for every state. Only one of the two formally degenerate eigenstates is physical.

\subsubsection*{Category 2}

A category-2 order parameter is generally a pair-density wave given by:
\begin{equation}\label{Cat2OP}
\eta_i = \eta_+ e^{i{\bf Q}{\bf r}_i} + \eta_- e^{-i{\bf Q}{\bf r}_i} \ ,
\end{equation}
where $\eta_\pm = (\eta_{\uparrow\pm}, \eta_{0\pm}, \eta_{\downarrow\pm})$ are three-component complex spinors and ${\bf Q}$ is a commensurate wavevector in the 1st Brillouin zone (1BZ) of the crystal lattice. In order to construct a BdG representation of the Hamiltonian, we must first identify the smallest positive integer $q$ for which $q{\bf Q}$ is a reciprocal lattice vector; $q$ specifies the size of a unit-cell in lattice sites for this periodic order parameter. Triplet pairing will mix fermion states at momenta separated by $\pm {\bf Q}$. In order to capture this, we define the Nambu spinor block in momentum space:
\begin{eqnarray}
&& \!\!\!\!\!\!\! \Psi_{\bf p}^{T}=\Bigl(\begin{array}{cccc}
\psi_{{\bf p}\uparrow+}^{\phantom{*}} & \psi_{(-{\bf p})\downarrow+}^{*} & \psi_{{\bf p}\downarrow+}^{\phantom{*}} & \psi_{(-{\bf p})\uparrow+}^{*} \end{array} \nonumber \\
&& ~~~~~~~ \begin{array}{cccc}  \psi_{{\bf p}\uparrow-}^{\phantom{*}} & \psi_{(-{\bf p})\downarrow-}^{*} & \psi_{{\bf p}\downarrow-}^{\phantom{*}} & \psi_{(-{\bf p})\uparrow-}^{*}\end{array}\Bigr) \nonumber \ ,
\end{eqnarray}
and divide the 1BZ zone into $q$ equal segments so that ${\bf p} = n{\bf Q} + {\bf k}$ with $n=0,\dots,q-1$. The residual wavevector ${\bf k}$ should live inside a reduced Brillouin zone whose area is $q$ times smaller than that of 1BZ. However, the shape of the reduced zone is restricted by ${\bf Q}$ in somewhat complicated ways. It is more practical to allow ${\bf k}$ to vary inside the full 1BZ and then divide any 1BZ integrals (e.g. in the calculation of the ground state energy) by $q$ to undo the overcounting of the BdG Hamiltonian eigenvalues.

The BdG Hamiltonian is again the Fourier transform of all fermion terms in (\ref{H0}) and (\ref{Hint}):
\begin{eqnarray}\label{HBdG3}
H_{\textrm{BdG}} \!\!&=&\!\! \int\limits_{\textrm{1BZ}} \frac{d^2 k}{(2\pi)^2} \sum_{nn'} \Psi_{n{\bf Q}+{\bf k}}^\dagger
  \Bigl\lbrack h_{n,{\bf k}} \delta_{nn'} \\
&& + h_{+}\delta_{n,n'+1}+h_{-}\delta_{n,n'-1} \Bigr\rbrack \Psi_{n'{\bf Q}+{\bf k}}^{\phantom{\dagger}} \ , \nonumber
\end{eqnarray}
where:
\begin{equation}
h_\pm = \left(\begin{array}{cccccccc}
0 & 0 & 0 & 0 & 0 & -\frac{\eta_{0\pm}^{\phantom{*}}}{\sqrt{2}} & 0 & -\eta_{\uparrow\pm}^{\phantom{*}}\\
0 & 0 & 0 & 0 & \frac{\eta_{0\mp}^{*}}{\sqrt{2}} & 0 & \eta_{\downarrow\mp}^{*} & 0\\
0 & 0 & 0 & 0 & 0 & -\eta_{\downarrow\pm}^{\phantom{*}} & 0 & -\frac{\eta_{0\pm}^{\phantom{*}}}{\sqrt{2}}\\
0 & 0 & 0 & 0 & \eta_{\uparrow\mp}^{*} & 0 & \frac{\eta_{0\mp}^{*}}{\sqrt{2}} & 0\\
0 & \frac{\eta_{0\pm}^{\phantom{*}}}{\sqrt{2}} & 0 & \eta_{\uparrow\pm}^{\phantom{*}} & 0 & 0 & 0 & 0\\
-\frac{\eta_{0\mp}^{*}}{\sqrt{2}} & 0 & -\eta_{\downarrow\mp}^{*} & 0 & 0 & 0 & 0 & 0\\
0 & \eta_{\downarrow\pm}^{\phantom{*}} & 0 & \frac{\eta_{0\pm}^{\phantom{*}}}{\sqrt{2}} & 0 & 0 & 0 & 0\\
-\eta_{\uparrow\mp}^{*} & 0 & -\frac{\eta_{0\mp}^{*}}{\sqrt{2}} & 0 & 0 & 0 & 0 & 0
\end{array}\right) \nonumber
\end{equation}
and
\begin{widetext}
\begin{equation}
h_{n,{\bf k}}=\left(\begin{array}{cccccccc}
E_{n{\bf Q}+{\bf k}} & -\phi_+ & V_{n{\bf Q}+{\bf k}} & 0 & \Delta & -\frac{1}{\sqrt{2}}\phi_0 & 0 & 0\\
-\phi_+^* & -E_{n{\bf Q}+{\bf k}} & 0 & V_{n{\bf Q}+{\bf k}} & -\frac{1}{\sqrt{2}}\phi_0^* & -\Delta & 0 & 0\\
V_{n{\bf Q}+{\bf k}}^{*} & 0 & E_{n{\bf Q}+{\bf k}} & \phi_+ & 0 & 0 & \Delta & \frac{1}{\sqrt{2}}\phi_0\\
0 & V_{n{\bf Q}+{\bf k}}^{*} & \phi_+^* & -E_{n{\bf Q}+{\bf k}} & 0 & 0 & \frac{1}{\sqrt{2}}\phi_0^* & -\Delta\\
\Delta & -\frac{1}{\sqrt{2}}\phi_0 & 0 & 0 & E_{n{\bf Q}+{\bf k}} & -\phi_- & -V_{n{\bf Q}+{\bf k}} & 0\\
-\frac{1}{\sqrt{2}}\phi_0^* & -\Delta & 0 & 0 & -\phi_-^* & -E_{n{\bf Q}+{\bf k}} & 0 & -V_{n{\bf Q}+{\bf k}}\\
0 & 0 & \Delta & \frac{1}{\sqrt{2}}\phi_0 & -V_{n{\bf Q}+{\bf k}}^{*} & 0 & E_{n{\bf Q}+{\bf k}} & \phi_-\\
0 & 0 & \frac{1}{\sqrt{2}}\phi_0^* & -\Delta & 0 & -V_{n{\bf Q}+{\bf k}}^{*} & \phi_-^* & -E_{n{\bf Q}+{\bf k}}
\end{array}\right) \nonumber
\end{equation}
\end{widetext}
\begin{equation}
\nonumber
\end{equation}
\begin{equation}
\nonumber
\end{equation}
with
\begin{eqnarray}\label{EV}
E_{\bf p} \!\!&=&\!\! -2t \Bigl\lbrack\cos(p_{x})+\cos(p_{y})\Bigr\rbrack \cos(a) - \mu \\
V_{\bf p} \!\!&=&\!\! 2t\Bigl\lbrack\sin(p_{y})+i\sin(p_{x})\Bigr\rbrack \sin(a) \ . \nonumber
\end{eqnarray}
We used the properties $E_{-{\bf p}} = E_{\bf p}$ and $V_{-{\bf p}} = -V_{\bf p}$ to construct the Hamiltonian. Once (\ref{HBdG3}) is diagonalized also with respect to the index $n$, one obtains the full set of Bloch states for the PDW order parameter.

\subsubsection*{Category 3}

The category-3 order parameter is a plane wave:
\begin{equation}\label{Cat3OP}
\eta_i = \eta_{+} e^{i{\bf Q}{\bf r}_i} \ .
\end{equation}
which is a subset of the category-2 order parameters (\ref{Cat2OP}) with $\eta_- = 0$ for commensurate ${\bf Q}$. However, since the triplet now lives at a single wavevector, we can diagonalize the BdG Hamiltonian even when ${\bf Q}$ is incommensurate. To that end, we define the Nambu spinor in momentum space as:
\begin{eqnarray}
&& \!\!\!\!\!\!\! \Psi_{\bf k}^{T}=\Bigl(\begin{array}{cccc}
\psi_{{\bf k}\uparrow+}^{\phantom{*}} & \psi_{({\bf Q}-{\bf k})\downarrow+}^{*} & \psi_{{\bf k}\downarrow+}^{\phantom{*}} &
  \psi_{({\bf Q}-{\bf k})\uparrow+}^{*} \end{array} \nonumber \\
&& ~~~~~~~ \begin{array}{cccc}  \psi_{{\bf k}\uparrow-}^{\phantom{*}} & \psi_{({\bf Q}-{\bf k})\downarrow-}^{*} &
  \psi_{{\bf k}\downarrow-}^{\phantom{*}} & \psi_{({\bf Q}-{\bf k})\uparrow-}^{*}\end{array}\Bigr) \nonumber
\end{eqnarray}
for every wavevector ${\bf k}$ in the full 1BZ of the crystal lattice. The BdG Hamiltonian is readily constructed in the same manner as for a uniform order parameter, but with a doubled representation that can accommodate triplet pairing. If we introduce the following symbols based on (\ref{EV}):
\begin{equation}
E_1 = E_{\bf k} \quad,\quad E_2 = E_{{\bf Q}-{\bf k}} \quad,\quad
V_1 = V_{\bf k} \quad,\quad V_2 = V_{{\bf Q}-{\bf k}} \nonumber
\end{equation}
then the BdG Hamiltonian takes the form:
\begin{equation}
H_{\textrm{BdG}} = \int\limits_{\textrm{1BZ}} \frac{d^2 k}{(2\pi)^2} \Psi_{\bf k}^\dagger h_{\bf k}^{\phantom{\dagger}} 
  \Psi_{\bf k}^{\phantom{\dagger}} \ , \nonumber
\end{equation}
where:
\bigskip
\begin{equation}
h_{\bf k}\!=\!\left(\!\!\begin{array}{cccccccc}
E_{1} & 0 & V_{1} & 0 & \Delta & -\frac{\eta_{0+}}{\sqrt{2}} & 0 & -\eta_{\uparrow+}\\
0 & -E_{2} & 0 & -V_{2} & \frac{\eta_{0+}^{*}}{\sqrt{2}} & -\Delta & \eta_{\downarrow+}^{*} & 0\\
V_{1}^{*} & 0 & E_{1} & 0 & 0 & -\eta_{\downarrow+} & \Delta & -\frac{\eta_{0+}}{\sqrt{2}}\\
0 & -V_{2}^{*} & 0 & -E_{2} & \eta_{\uparrow+}^{*} & 0 & \frac{\eta_{0+}^{*}}{\sqrt{2}} & -\Delta\\
\Delta & \frac{\eta_{0+}}{\sqrt{2}} & 0 & \eta_{\uparrow+} & E_{1} & 0 & -V_{1} & 0\\
-\frac{\eta_{0+}^{*}}{\sqrt{2}} & -\Delta & -\eta_{\downarrow+}^{*} & 0 & 0 & -E_{2} & 0 & V_{2}\\
0 & \eta_{\downarrow+} & \Delta & \frac{\eta_{0+}}{\sqrt{2}} & -V_{1}^{*} & 0 & E_{1} & 0\\
-\eta_{\uparrow+}^{*} & 0 & -\frac{\eta_{0+}^{*}}{\sqrt{2}} & -\Delta & 0 & V_{2}^{*} & 0 & -E_{2}
\end{array}\!\!\right) \nonumber
\end{equation}
Note that there is no room for uniform singlet pairing in this representation, unless ${\bf Q}=0$.


\begin{thebibliography}{41}%
\makeatletter
\providecommand \@ifxundefined [1]{%
 \@ifx{#1\undefined}
}%
\providecommand \@ifnum [1]{%
 \ifnum #1\expandafter \@firstoftwo
 \else \expandafter \@secondoftwo
 \fi
}%
\providecommand \@ifx [1]{%
 \ifx #1\expandafter \@firstoftwo
 \else \expandafter \@secondoftwo
 \fi
}%
\providecommand \natexlab [1]{#1}%
\providecommand \enquote  [1]{``#1''}%
\providecommand \bibnamefont  [1]{#1}%
\providecommand \bibfnamefont [1]{#1}%
\providecommand \citenamefont [1]{#1}%
\providecommand \href@noop [0]{\@secondoftwo}%
\providecommand \href [0]{\begingroup \@sanitize@url \@href}%
\providecommand \@href[1]{\@@startlink{#1}\@@href}%
\providecommand \@@href[1]{\endgroup#1\@@endlink}%
\providecommand \@sanitize@url [0]{\catcode `\\12\catcode `\$12\catcode
  `\&12\catcode `\#12\catcode `\^12\catcode `\_12\catcode `\%12\relax}%
\providecommand \@@startlink[1]{}%
\providecommand \@@endlink[0]{}%
\providecommand \url  [0]{\begingroup\@sanitize@url \@url }%
\providecommand \@url [1]{\endgroup\@href {#1}{\urlprefix }}%
\providecommand \urlprefix  [0]{URL }%
\providecommand \Eprint [0]{\href }%
\providecommand \doibase [0]{http://dx.doi.org/}%
\providecommand \selectlanguage [0]{\@gobble}%
\providecommand \bibinfo  [0]{\@secondoftwo}%
\providecommand \bibfield  [0]{\@secondoftwo}%
\providecommand \translation [1]{[#1]}%
\providecommand \BibitemOpen [0]{}%
\providecommand \bibitemStop [0]{}%
\providecommand \bibitemNoStop [0]{.\EOS\space}%
\providecommand \EOS [0]{\spacefactor3000\relax}%
\providecommand \BibitemShut  [1]{\csname bibitem#1\endcsname}%
\let\auto@bib@innerbib\@empty
\bibitem [{\citenamefont {Fr{\"o}hlich}\ and\ \citenamefont
  {Studer}(1992)}]{Frohlich1992}%
  \BibitemOpen
  \bibfield  {author} {\bibinfo {author} {\bibfnamefont {J.}~\bibnamefont
  {Fr{\"o}hlich}}\ and\ \bibinfo {author} {\bibfnamefont {U.~M.}\ \bibnamefont
  {Studer}},\ }\href@noop {} {\bibfield  {journal} {\bibinfo  {journal}
  {Communications in Mathematical Physics}\ }\textbf {\bibinfo {volume}
  {148}},\ \bibinfo {pages} {553} (\bibinfo {year} {1992})}\BibitemShut
  {NoStop}%
\bibitem [{\citenamefont {Nikolic}\ \emph {et~al.}(2013)\citenamefont
  {Nikolic}, \citenamefont {Duric},\ and\ \citenamefont
  {Tesanovic}}]{Nikolic2011a}%
  \BibitemOpen
  \bibfield  {author} {\bibinfo {author} {\bibfnamefont {P.}~\bibnamefont
  {Nikolic}}, \bibinfo {author} {\bibfnamefont {T.}~\bibnamefont {Duric}}, \
  and\ \bibinfo {author} {\bibfnamefont {Z.}~\bibnamefont {Tesanovic}},\
  }\href@noop {} {\bibfield  {journal} {\bibinfo  {journal} {Physical Review
  Letters}\ }\textbf {\bibinfo {volume} {110}},\ \bibinfo {pages} {176804}
  (\bibinfo {year} {2013})}\BibitemShut {NoStop}%
\bibitem [{\citenamefont {Murakami}\ \emph {et~al.}(2004)\citenamefont
  {Murakami}, \citenamefont {Nagaosa},\ and\ \citenamefont
  {Zhang}}]{Murakami2004}%
  \BibitemOpen
  \bibfield  {author} {\bibinfo {author} {\bibfnamefont {S.}~\bibnamefont
  {Murakami}}, \bibinfo {author} {\bibfnamefont {N.}~\bibnamefont {Nagaosa}}, \
  and\ \bibinfo {author} {\bibfnamefont {S.-C.}\ \bibnamefont {Zhang}},\
  }\href@noop {} {\bibfield  {journal} {\bibinfo  {journal} {Physical Review
  Letters}\ }\textbf {\bibinfo {volume} {93}},\ \bibinfo {pages} {156804}
  (\bibinfo {year} {2004})}\BibitemShut {NoStop}%
\bibitem [{\citenamefont {Kane}\ and\ \citenamefont {Mele}(2005)}]{Kane2005}%
  \BibitemOpen
  \bibfield  {author} {\bibinfo {author} {\bibfnamefont {C.~L.}\ \bibnamefont
  {Kane}}\ and\ \bibinfo {author} {\bibfnamefont {E.~J.}\ \bibnamefont
  {Mele}},\ }\href@noop {} {\bibfield  {journal} {\bibinfo  {journal} {Physical
  Review Letters}\ }\textbf {\bibinfo {volume} {95}},\ \bibinfo {pages}
  {226801} (\bibinfo {year} {2005})}\BibitemShut {NoStop}%
\bibitem [{\citenamefont {Bernevig}\ \emph {et~al.}(2006)\citenamefont
  {Bernevig}, \citenamefont {Hughes},\ and\ \citenamefont
  {Zhang}}]{Bernevig2006}%
  \BibitemOpen
  \bibfield  {author} {\bibinfo {author} {\bibfnamefont {B.~A.}\ \bibnamefont
  {Bernevig}}, \bibinfo {author} {\bibfnamefont {T.~L.}\ \bibnamefont
  {Hughes}}, \ and\ \bibinfo {author} {\bibfnamefont {S.-C.}\ \bibnamefont
  {Zhang}},\ }\href@noop {} {\bibfield  {journal} {\bibinfo  {journal}
  {Science}\ }\textbf {\bibinfo {volume} {314}},\ \bibinfo {pages} {1757}
  (\bibinfo {year} {2006})}\BibitemShut {NoStop}%
\bibitem [{\citenamefont {K{\"o}nig}\ \emph {et~al.}(2007)\citenamefont
  {K{\"o}nig}, \citenamefont {Wiedmann}, \citenamefont {Br{\"u}ne},
  \citenamefont {Roth}, \citenamefont {Buhmann}, \citenamefont {Molenkamp},
  \citenamefont {Qi},\ and\ \citenamefont {Zhang}}]{Konig2007}%
  \BibitemOpen
  \bibfield  {author} {\bibinfo {author} {\bibfnamefont {M.}~\bibnamefont
  {K{\"o}nig}}, \bibinfo {author} {\bibfnamefont {S.}~\bibnamefont {Wiedmann}},
  \bibinfo {author} {\bibfnamefont {C.}~\bibnamefont {Br{\"u}ne}}, \bibinfo
  {author} {\bibfnamefont {A.}~\bibnamefont {Roth}}, \bibinfo {author}
  {\bibfnamefont {H.}~\bibnamefont {Buhmann}}, \bibinfo {author} {\bibfnamefont
  {L.~W.}\ \bibnamefont {Molenkamp}}, \bibinfo {author} {\bibfnamefont {X.-L.}\
  \bibnamefont {Qi}}, \ and\ \bibinfo {author} {\bibfnamefont {S.-C.}\
  \bibnamefont {Zhang}},\ }\href@noop {} {\bibfield  {journal} {\bibinfo
  {journal} {Science}\ }\textbf {\bibinfo {volume} {318}},\ \bibinfo {pages}
  {766} (\bibinfo {year} {2007})}\BibitemShut {NoStop}%
\bibitem [{\citenamefont {Spielman}(2009)}]{Spielman2009}%
  \BibitemOpen
  \bibfield  {author} {\bibinfo {author} {\bibfnamefont {I.~B.}\ \bibnamefont
  {Spielman}},\ }\href@noop {} {\bibfield  {journal} {\bibinfo  {journal}
  {Physical Review A}\ }\textbf {\bibinfo {volume} {79}},\ \bibinfo {pages}
  {063613} (\bibinfo {year} {2009})}\BibitemShut {NoStop}%
\bibitem [{\citenamefont {Lin}\ \emph {et~al.}(2011)\citenamefont {Lin},
  \citenamefont {Jim{\'e}nez-Garc{\'i}a},\ and\ \citenamefont
  {Spielman}}]{Lin2011}%
  \BibitemOpen
  \bibfield  {author} {\bibinfo {author} {\bibfnamefont {Y.-J.}\ \bibnamefont
  {Lin}}, \bibinfo {author} {\bibfnamefont {K.}~\bibnamefont
  {Jim{\'e}nez-Garc{\'i}a}}, \ and\ \bibinfo {author} {\bibfnamefont {I.~B.}\
  \bibnamefont {Spielman}},\ }\href@noop {} {\bibfield  {journal} {\bibinfo
  {journal} {Nature}\ }\textbf {\bibinfo {volume} {471}},\ \bibinfo {pages}
  {83} (\bibinfo {year} {2011})}\BibitemShut {NoStop}%
\bibitem [{\citenamefont {Wang}\ \emph {et~al.}(2012)\citenamefont {Wang},
  \citenamefont {Yu}, \citenamefont {Fu}, \citenamefont {Miao}, \citenamefont
  {Huang}, \citenamefont {Chai}, \citenamefont {Zhai},\ and\ \citenamefont
  {Zhang}}]{Wang2012b}%
  \BibitemOpen
  \bibfield  {author} {\bibinfo {author} {\bibfnamefont {P.}~\bibnamefont
  {Wang}}, \bibinfo {author} {\bibfnamefont {Z.-Q.}\ \bibnamefont {Yu}},
  \bibinfo {author} {\bibfnamefont {Z.}~\bibnamefont {Fu}}, \bibinfo {author}
  {\bibfnamefont {J.}~\bibnamefont {Miao}}, \bibinfo {author} {\bibfnamefont
  {L.}~\bibnamefont {Huang}}, \bibinfo {author} {\bibfnamefont
  {S.}~\bibnamefont {Chai}}, \bibinfo {author} {\bibfnamefont {H.}~\bibnamefont
  {Zhai}}, \ and\ \bibinfo {author} {\bibfnamefont {J.}~\bibnamefont {Zhang}},\
  }\href@noop {} {\bibfield  {journal} {\bibinfo  {journal} {Physical Review
  Letters}\ }\textbf {\bibinfo {volume} {109}},\ \bibinfo {pages} {095301}
  (\bibinfo {year} {2012})}\BibitemShut {NoStop}%
\bibitem [{\citenamefont {Cheuk}\ \emph {et~al.}(2012)\citenamefont {Cheuk},
  \citenamefont {Sommer}, \citenamefont {Hadzibabic}, \citenamefont {Yefsah},
  \citenamefont {Bakr},\ and\ \citenamefont {Zwierlein}}]{Cheuk2012}%
  \BibitemOpen
  \bibfield  {author} {\bibinfo {author} {\bibfnamefont {L.~W.}\ \bibnamefont
  {Cheuk}}, \bibinfo {author} {\bibfnamefont {A.~T.}\ \bibnamefont {Sommer}},
  \bibinfo {author} {\bibfnamefont {Z.}~\bibnamefont {Hadzibabic}}, \bibinfo
  {author} {\bibfnamefont {T.}~\bibnamefont {Yefsah}}, \bibinfo {author}
  {\bibfnamefont {W.~S.}\ \bibnamefont {Bakr}}, \ and\ \bibinfo {author}
  {\bibfnamefont {M.~W.}\ \bibnamefont {Zwierlein}},\ }\href@noop {} {\bibfield
   {journal} {\bibinfo  {journal} {Physical Review Letters}\ }\textbf {\bibinfo
  {volume} {109}},\ \bibinfo {pages} {095302} (\bibinfo {year}
  {2012})}\BibitemShut {NoStop}%
\bibitem [{\citenamefont {Ho}\ and\ \citenamefont {Zhang}(2011)}]{Ho2011}%
  \BibitemOpen
  \bibfield  {author} {\bibinfo {author} {\bibfnamefont {T.-L.}\ \bibnamefont
  {Ho}}\ and\ \bibinfo {author} {\bibfnamefont {S.}~\bibnamefont {Zhang}},\
  }\href@noop {} {\bibfield  {journal} {\bibinfo  {journal} {Physical Review
  Letters}\ }\textbf {\bibinfo {volume} {107}},\ \bibinfo {pages} {150403}
  (\bibinfo {year} {2011})}\BibitemShut {NoStop}%
\bibitem [{\citenamefont {Wang}\ \emph {et~al.}(2010)\citenamefont {Wang},
  \citenamefont {Gao}, \citenamefont {Jian},\ and\ \citenamefont
  {Zhai}}]{Wang2010b}%
  \BibitemOpen
  \bibfield  {author} {\bibinfo {author} {\bibfnamefont {C.}~\bibnamefont
  {Wang}}, \bibinfo {author} {\bibfnamefont {C.}~\bibnamefont {Gao}}, \bibinfo
  {author} {\bibfnamefont {C.-M.}\ \bibnamefont {Jian}}, \ and\ \bibinfo
  {author} {\bibfnamefont {H.}~\bibnamefont {Zhai}},\ }\href@noop {} {\bibfield
   {journal} {\bibinfo  {journal} {Physical Review Letters}\ }\textbf {\bibinfo
  {volume} {105}},\ \bibinfo {pages} {160403} (\bibinfo {year}
  {2010})}\BibitemShut {NoStop}%
\bibitem [{\citenamefont {Zhai}(2012)}]{Zhai2012}%
  \BibitemOpen
  \bibfield  {author} {\bibinfo {author} {\bibfnamefont {H.}~\bibnamefont
  {Zhai}},\ }\href@noop {} {\bibfield  {journal} {\bibinfo  {journal}
  {International Journal of Modern Physics B}\ }\textbf {\bibinfo {volume}
  {26}},\ \bibinfo {pages} {1230001} (\bibinfo {year} {2012})}\BibitemShut
  {NoStop}%
\bibitem [{\citenamefont {Ozawa}\ and\ \citenamefont {Baym}(2012)}]{Ozawa2012}%
  \BibitemOpen
  \bibfield  {author} {\bibinfo {author} {\bibfnamefont {T.}~\bibnamefont
  {Ozawa}}\ and\ \bibinfo {author} {\bibfnamefont {G.}~\bibnamefont {Baym}},\
  }\href@noop {} {\bibfield  {journal} {\bibinfo  {journal} {Physical Review
  A}\ }\textbf {\bibinfo {volume} {85}},\ \bibinfo {pages} {013612} (\bibinfo
  {year} {2012})}\BibitemShut {NoStop}%
\bibitem [{\citenamefont {Sedrakyan}\ \emph {et~al.}(2012)\citenamefont
  {Sedrakyan}, \citenamefont {Kamenev},\ and\ \citenamefont
  {Glazman}}]{Sedrakyan2012}%
  \BibitemOpen
  \bibfield  {author} {\bibinfo {author} {\bibfnamefont {T.~A.}\ \bibnamefont
  {Sedrakyan}}, \bibinfo {author} {\bibfnamefont {A.}~\bibnamefont {Kamenev}},
  \ and\ \bibinfo {author} {\bibfnamefont {L.~I.}\ \bibnamefont {Glazman}},\
  }\href@noop {} {\bibfield  {journal} {\bibinfo  {journal} {Physical Review
  A}\ }\textbf {\bibinfo {volume} {86}},\ \bibinfo {pages} {063639} (\bibinfo
  {year} {2012})}\BibitemShut {NoStop}%
\bibitem [{\citenamefont {Nikolic}\ and\ \citenamefont
  {Tesanovic}(2013{\natexlab{a}})}]{Nikolic2012b}%
  \BibitemOpen
  \bibfield  {author} {\bibinfo {author} {\bibfnamefont {P.}~\bibnamefont
  {Nikolic}}\ and\ \bibinfo {author} {\bibfnamefont {Z.}~\bibnamefont
  {Tesanovic}},\ }\href@noop {} {\bibfield  {journal} {\bibinfo  {journal}
  {Physical Review B}\ }\textbf {\bibinfo {volume} {87}},\ \bibinfo {pages}
  {134511} (\bibinfo {year} {2013}{\natexlab{a}})}\BibitemShut {NoStop}%
\bibitem [{\citenamefont {Nikoli{\'c}}(2014{\natexlab{a}})}]{Nikolic2014b}%
  \BibitemOpen
  \bibfield  {author} {\bibinfo {author} {\bibfnamefont {P.}~\bibnamefont
  {Nikoli{\'c}}},\ }\href@noop {} {\bibfield  {journal} {\bibinfo  {journal}
  {Physical Review B}\ }\textbf {\bibinfo {volume} {90}},\ \bibinfo {pages}
  {235107} (\bibinfo {year} {2014}{\natexlab{a}})}\BibitemShut {NoStop}%
\bibitem [{\citenamefont {Nikoli{\'c}}(2014{\natexlab{b}})}]{Nikolic2014}%
  \BibitemOpen
  \bibfield  {author} {\bibinfo {author} {\bibfnamefont {P.}~\bibnamefont
  {Nikoli{\'c}}},\ }\href@noop {} {\bibfield  {journal} {\bibinfo  {journal}
  {Physical Review A}\ }\textbf {\bibinfo {volume} {90}},\ \bibinfo {pages}
  {023623} (\bibinfo {year} {2014}{\natexlab{b}})}\BibitemShut {NoStop}%
\bibitem [{\citenamefont {Cole}\ \emph {et~al.}(2012)\citenamefont {Cole},
  \citenamefont {Zhang}, \citenamefont {Paramekanti},\ and\ \citenamefont
  {Trivedi}}]{Cole2012}%
  \BibitemOpen
  \bibfield  {author} {\bibinfo {author} {\bibfnamefont {W.~S.}\ \bibnamefont
  {Cole}}, \bibinfo {author} {\bibfnamefont {S.}~\bibnamefont {Zhang}},
  \bibinfo {author} {\bibfnamefont {A.}~\bibnamefont {Paramekanti}}, \ and\
  \bibinfo {author} {\bibfnamefont {N.}~\bibnamefont {Trivedi}},\ }\href@noop
  {} {\bibfield  {journal} {\bibinfo  {journal} {Physical Review Letters}\
  }\textbf {\bibinfo {volume} {109}},\ \bibinfo {pages} {085302} (\bibinfo
  {year} {2012})}\BibitemShut {NoStop}%
\bibitem [{\citenamefont {Radic}\ \emph {et~al.}(2012)\citenamefont {Radic},
  \citenamefont {Ciolo}, \citenamefont {Sun},\ and\ \citenamefont
  {Galitski}}]{Radic2012}%
  \BibitemOpen
  \bibfield  {author} {\bibinfo {author} {\bibfnamefont {J.}~\bibnamefont
  {Radic}}, \bibinfo {author} {\bibfnamefont {A.~D.}\ \bibnamefont {Ciolo}},
  \bibinfo {author} {\bibfnamefont {K.}~\bibnamefont {Sun}}, \ and\ \bibinfo
  {author} {\bibfnamefont {V.}~\bibnamefont {Galitski}},\ }\href@noop {}
  {\bibfield  {journal} {\bibinfo  {journal} {Physical Review Letters}\
  }\textbf {\bibinfo {volume} {109}},\ \bibinfo {pages} {085303} (\bibinfo
  {year} {2012})}\BibitemShut {NoStop}%
\bibitem [{\citenamefont {Cai}\ \emph {et~al.}(2012)\citenamefont {Cai},
  \citenamefont {Zhou},\ and\ \citenamefont {Wu}}]{Cai2012}%
  \BibitemOpen
  \bibfield  {author} {\bibinfo {author} {\bibfnamefont {Z.}~\bibnamefont
  {Cai}}, \bibinfo {author} {\bibfnamefont {X.}~\bibnamefont {Zhou}}, \ and\
  \bibinfo {author} {\bibfnamefont {C.}~\bibnamefont {Wu}},\ }\href@noop {}
  {\bibfield  {journal} {\bibinfo  {journal} {Physical Review A}\ }\textbf
  {\bibinfo {volume} {85}},\ \bibinfo {pages} {061605(R)} (\bibinfo {year}
  {2012})}\BibitemShut {NoStop}%
\bibitem [{\citenamefont {Nikolic}(2009)}]{nikolic:144507}%
  \BibitemOpen
  \bibfield  {author} {\bibinfo {author} {\bibfnamefont {P.}~\bibnamefont
  {Nikolic}},\ }\href@noop {} {\bibfield  {journal} {\bibinfo  {journal}
  {Physical Review B}\ }\textbf {\bibinfo {volume} {79}},\ \bibinfo {pages}
  {144507} (\bibinfo {year} {2009})}\BibitemShut {NoStop}%
\bibitem [{\citenamefont {Wilkin}\ and\ \citenamefont
  {Gunn}(2000)}]{Wilkin2000}%
  \BibitemOpen
  \bibfield  {author} {\bibinfo {author} {\bibfnamefont {N.~K.}\ \bibnamefont
  {Wilkin}}\ and\ \bibinfo {author} {\bibfnamefont {J.~M.~F.}\ \bibnamefont
  {Gunn}},\ }\href@noop {} {\bibfield  {journal} {\bibinfo  {journal} {Physical
  Review Letters}\ }\textbf {\bibinfo {volume} {84}},\ \bibinfo {pages} {6}
  (\bibinfo {year} {2000})}\BibitemShut {NoStop}%
\bibitem [{\citenamefont {Cooper}\ \emph {et~al.}(2001)\citenamefont {Cooper},
  \citenamefont {Wilkin},\ and\ \citenamefont {Gunn}}]{Cooper2001}%
  \BibitemOpen
  \bibfield  {author} {\bibinfo {author} {\bibfnamefont {N.~R.}\ \bibnamefont
  {Cooper}}, \bibinfo {author} {\bibfnamefont {N.~K.}\ \bibnamefont {Wilkin}},
  \ and\ \bibinfo {author} {\bibfnamefont {J.~M.~F.}\ \bibnamefont {Gunn}},\
  }\href@noop {} {\bibfield  {journal} {\bibinfo  {journal} {Physical Review
  Letters}\ }\textbf {\bibinfo {volume} {87}},\ \bibinfo {pages} {120405}
  (\bibinfo {year} {2001})}\BibitemShut {NoStop}%
\bibitem [{\citenamefont {Regnault}\ and\ \citenamefont
  {Jolicoeur}(2003)}]{Regnault2003}%
  \BibitemOpen
  \bibfield  {author} {\bibinfo {author} {\bibfnamefont {N.}~\bibnamefont
  {Regnault}}\ and\ \bibinfo {author} {\bibfnamefont {T.}~\bibnamefont
  {Jolicoeur}},\ }\href@noop {} {\bibfield  {journal} {\bibinfo  {journal}
  {Physical Review Letters}\ }\textbf {\bibinfo {volume} {91}},\ \bibinfo
  {pages} {030402} (\bibinfo {year} {2003})}\BibitemShut {NoStop}%
\bibitem [{\citenamefont {Chang}\ \emph {et~al.}(2005)\citenamefont {Chang},
  \citenamefont {Regnault}, \citenamefont {Jolicoeur},\ and\ \citenamefont
  {Jain}}]{Chang2005}%
  \BibitemOpen
  \bibfield  {author} {\bibinfo {author} {\bibfnamefont {C.~C.}\ \bibnamefont
  {Chang}}, \bibinfo {author} {\bibfnamefont {N.}~\bibnamefont {Regnault}},
  \bibinfo {author} {\bibfnamefont {T.}~\bibnamefont {Jolicoeur}}, \ and\
  \bibinfo {author} {\bibfnamefont {J.~K.}\ \bibnamefont {Jain}},\ }\href@noop
  {} {\bibfield  {journal} {\bibinfo  {journal} {Physical Review A}\ }\textbf
  {\bibinfo {volume} {72}},\ \bibinfo {pages} {013611} (\bibinfo {year}
  {2005})}\BibitemShut {NoStop}%
\bibitem [{\citenamefont {Cooper}(2008)}]{Cooper2008}%
  \BibitemOpen
  \bibfield  {author} {\bibinfo {author} {\bibfnamefont {N.~R.}\ \bibnamefont
  {Cooper}},\ }\href@noop {} {\bibfield  {journal} {\bibinfo  {journal}
  {Advances in Physics}\ }\textbf {\bibinfo {volume} {57}},\ \bibinfo {pages}
  {539} (\bibinfo {year} {2008})}\BibitemShut {NoStop}%
\bibitem [{\citenamefont {Nikolic}(2013{\natexlab{a}})}]{Nikolic2011}%
  \BibitemOpen
  \bibfield  {author} {\bibinfo {author} {\bibfnamefont {P.}~\bibnamefont
  {Nikolic}},\ }\href@noop {} {\bibfield  {journal} {\bibinfo  {journal}
  {Journal of Physics: Condensed Matter}\ }\textbf {\bibinfo {volume} {25}},\
  \bibinfo {pages} {025602} (\bibinfo {year} {2013}{\natexlab{a}})}\BibitemShut
  {NoStop}%
\bibitem [{\citenamefont {Nikolic}(2013{\natexlab{b}})}]{Nikolic2012}%
  \BibitemOpen
  \bibfield  {author} {\bibinfo {author} {\bibfnamefont {P.}~\bibnamefont
  {Nikolic}},\ }\href@noop {} {\bibfield  {journal} {\bibinfo  {journal}
  {Physical Review B}\ }\textbf {\bibinfo {volume} {87}},\ \bibinfo {pages}
  {245120} (\bibinfo {year} {2013}{\natexlab{b}})}\BibitemShut {NoStop}%
\bibitem [{\citenamefont {Fu}\ \emph {et~al.}(2007)\citenamefont {Fu},
  \citenamefont {Kane},\ and\ \citenamefont {Mele}}]{Fu2007}%
  \BibitemOpen
  \bibfield  {author} {\bibinfo {author} {\bibfnamefont {L.}~\bibnamefont
  {Fu}}, \bibinfo {author} {\bibfnamefont {C.~L.}\ \bibnamefont {Kane}}, \ and\
  \bibinfo {author} {\bibfnamefont {E.~J.}\ \bibnamefont {Mele}},\ }\href@noop
  {} {\bibfield  {journal} {\bibinfo  {journal} {Physical Review Letters}\
  }\textbf {\bibinfo {volume} {98}},\ \bibinfo {pages} {106803} (\bibinfo
  {year} {2007})}\BibitemShut {NoStop}%
\bibitem [{\citenamefont {Nickerson}\ \emph {et~al.}(1971)\citenamefont
  {Nickerson}, \citenamefont {White}, \citenamefont {Lee}, \citenamefont
  {Bachmann}, \citenamefont {Geballe},\ and\ \citenamefont {{G. W.
  Hull}}}]{Nickerson1971}%
  \BibitemOpen
  \bibfield  {author} {\bibinfo {author} {\bibfnamefont {J.~C.}\ \bibnamefont
  {Nickerson}}, \bibinfo {author} {\bibfnamefont {R.~M.}\ \bibnamefont
  {White}}, \bibinfo {author} {\bibfnamefont {K.~N.}\ \bibnamefont {Lee}},
  \bibinfo {author} {\bibfnamefont {R.}~\bibnamefont {Bachmann}}, \bibinfo
  {author} {\bibfnamefont {T.~H.}\ \bibnamefont {Geballe}}, \ and\ \bibinfo
  {author} {\bibfnamefont {J.}~\bibnamefont {{G. W. Hull}}},\ }\href@noop {}
  {\bibfield  {journal} {\bibinfo  {journal} {Physical Review B}\ }\textbf
  {\bibinfo {volume} {3}},\ \bibinfo {pages} {2030} (\bibinfo {year}
  {1971})}\BibitemShut {NoStop}%
\bibitem [{\citenamefont {Farberovich}\ \emph {et~al.}(1983)\citenamefont
  {Farberovich}, \citenamefont {Kurganskii}, \citenamefont {Sidorin},
  \citenamefont {Karin}, \citenamefont {Bobrikov}, \citenamefont {Nizhnikova},
  \citenamefont {Shelikh}, \citenamefont {Korsukova},\ and\ \citenamefont
  {Gurin}}]{Farberovich1983}%
  \BibitemOpen
  \bibfield  {author} {\bibinfo {author} {\bibfnamefont {O.~V.}\ \bibnamefont
  {Farberovich}}, \bibinfo {author} {\bibfnamefont {S.~I.}\ \bibnamefont
  {Kurganskii}}, \bibinfo {author} {\bibfnamefont {K.~K.}\ \bibnamefont
  {Sidorin}}, \bibinfo {author} {\bibfnamefont {M.~G.}\ \bibnamefont {Karin}},
  \bibinfo {author} {\bibfnamefont {V.~N.}\ \bibnamefont {Bobrikov}}, \bibinfo
  {author} {\bibfnamefont {G.~P.}\ \bibnamefont {Nizhnikova}}, \bibinfo
  {author} {\bibnamefont {Shelikh}}, \bibinfo {author} {\bibfnamefont {M.~M.}\
  \bibnamefont {Korsukova}}, \ and\ \bibinfo {author} {\bibfnamefont {V.~N.}\
  \bibnamefont {Gurin}},\ }\href@noop {} {\bibfield  {journal} {\bibinfo
  {journal} {Sov. Phys. Solid. State}\ }\textbf {\bibinfo {volume} {25}},\
  \bibinfo {pages} {404} (\bibinfo {year} {1983})}\BibitemShut {NoStop}%
\bibitem [{\citenamefont {Takimoto}(2011)}]{Takimoto2011}%
  \BibitemOpen
  \bibfield  {author} {\bibinfo {author} {\bibfnamefont {T.}~\bibnamefont
  {Takimoto}},\ }\href@noop {} {\bibfield  {journal} {\bibinfo  {journal}
  {Journal of the Physical Society of Japan}\ }\textbf {\bibinfo {volume}
  {80}},\ \bibinfo {pages} {123710} (\bibinfo {year} {2011})}\BibitemShut
  {NoStop}%
\bibitem [{\citenamefont {Kang}\ \emph {et~al.}(2013)\citenamefont {Kang},
  \citenamefont {Kim}, \citenamefont {Kim}, \citenamefont {Kang}, \citenamefont
  {Denlinger},\ and\ \citenamefont {Min}}]{Kang2013}%
  \BibitemOpen
  \bibfield  {author} {\bibinfo {author} {\bibfnamefont {C.-J.}\ \bibnamefont
  {Kang}}, \bibinfo {author} {\bibfnamefont {J.}~\bibnamefont {Kim}}, \bibinfo
  {author} {\bibfnamefont {K.}~\bibnamefont {Kim}}, \bibinfo {author}
  {\bibfnamefont {J.-S.}\ \bibnamefont {Kang}}, \bibinfo {author}
  {\bibfnamefont {J.~D.}\ \bibnamefont {Denlinger}}, \ and\ \bibinfo {author}
  {\bibfnamefont {B.~I.}\ \bibnamefont {Min}},\ }\href@noop {} {\  (\bibinfo
  {year} {2013})},\ \bibinfo {note} {arXiv:1312.5898}\BibitemShut {NoStop}%
\bibitem [{\citenamefont {Lu}\ \emph {et~al.}(2013)\citenamefont {Lu},
  \citenamefont {Zhao}, \citenamefont {Weng}, \citenamefont {Fang},\ and\
  \citenamefont {Dai}}]{Lu2013b}%
  \BibitemOpen
  \bibfield  {author} {\bibinfo {author} {\bibfnamefont {F.}~\bibnamefont
  {Lu}}, \bibinfo {author} {\bibfnamefont {J.}~\bibnamefont {Zhao}}, \bibinfo
  {author} {\bibfnamefont {H.}~\bibnamefont {Weng}}, \bibinfo {author}
  {\bibfnamefont {Z.}~\bibnamefont {Fang}}, \ and\ \bibinfo {author}
  {\bibfnamefont {X.}~\bibnamefont {Dai}},\ }\href@noop {} {\bibfield
  {journal} {\bibinfo  {journal} {Physical Review Letters}\ }\textbf {\bibinfo
  {volume} {110}},\ \bibinfo {pages} {096401} (\bibinfo {year}
  {2013})}\BibitemShut {NoStop}%
\bibitem [{\citenamefont {Fuhrman}\ \emph {et~al.}(2015)\citenamefont
  {Fuhrman}, \citenamefont {Leiner}, \citenamefont {Nikoli{\'c}}, \citenamefont
  {Granroth}, \citenamefont {Stone}, \citenamefont {Lumsden}, \citenamefont
  {DeBeer-Schmitt}, \citenamefont {Alekseev}, \citenamefont {Mignot},
  \citenamefont {Koohpayeh}, \citenamefont {Cottingham}, \citenamefont
  {Phelan}, \citenamefont {Schoop}, \citenamefont {McQueen},\ and\
  \citenamefont {Broholm}}]{Fuhrman2014}%
  \BibitemOpen
  \bibfield  {author} {\bibinfo {author} {\bibfnamefont {W.~T.}\ \bibnamefont
  {Fuhrman}}, \bibinfo {author} {\bibfnamefont {J.}~\bibnamefont {Leiner}},
  \bibinfo {author} {\bibfnamefont {P.}~\bibnamefont {Nikoli{\'c}}}, \bibinfo
  {author} {\bibfnamefont {G.~E.}\ \bibnamefont {Granroth}}, \bibinfo {author}
  {\bibfnamefont {M.~B.}\ \bibnamefont {Stone}}, \bibinfo {author}
  {\bibfnamefont {M.~D.}\ \bibnamefont {Lumsden}}, \bibinfo {author}
  {\bibfnamefont {L.}~\bibnamefont {DeBeer-Schmitt}}, \bibinfo {author}
  {\bibfnamefont {P.~A.}\ \bibnamefont {Alekseev}}, \bibinfo {author}
  {\bibfnamefont {J.-M.}\ \bibnamefont {Mignot}}, \bibinfo {author}
  {\bibfnamefont {S.~M.}\ \bibnamefont {Koohpayeh}}, \bibinfo {author}
  {\bibfnamefont {P.}~\bibnamefont {Cottingham}}, \bibinfo {author}
  {\bibfnamefont {W.~A.}\ \bibnamefont {Phelan}}, \bibinfo {author}
  {\bibfnamefont {L.}~\bibnamefont {Schoop}}, \bibinfo {author} {\bibfnamefont
  {T.~M.}\ \bibnamefont {McQueen}}, \ and\ \bibinfo {author} {\bibfnamefont
  {C.}~\bibnamefont {Broholm}},\ }\href@noop {} {\bibfield  {journal} {\bibinfo
   {journal} {Physical Review Letters}\ }\textbf {\bibinfo {volume} {114}},\
  \bibinfo {pages} {036401} (\bibinfo {year} {2015})}\BibitemShut {NoStop}%
\bibitem [{\citenamefont {Fuhrman}\ and\ \citenamefont
  {Nikoli{\'c}}(2014)}]{Nikolic2014c}%
  \BibitemOpen
  \bibfield  {author} {\bibinfo {author} {\bibfnamefont {W.~T.}\ \bibnamefont
  {Fuhrman}}\ and\ \bibinfo {author} {\bibfnamefont {P.}~\bibnamefont
  {Nikoli{\'c}}},\ }\href@noop {} {\bibfield  {journal} {\bibinfo  {journal}
  {Physical Review B}\ }\textbf {\bibinfo {volume} {90}},\ \bibinfo {pages}
  {195144} (\bibinfo {year} {2014})}\BibitemShut {NoStop}%
\bibitem [{\citenamefont {Nikolic}\ and\ \citenamefont
  {Tesanovic}(2013{\natexlab{b}})}]{Nikolic2012a}%
  \BibitemOpen
  \bibfield  {author} {\bibinfo {author} {\bibfnamefont {P.}~\bibnamefont
  {Nikolic}}\ and\ \bibinfo {author} {\bibfnamefont {Z.}~\bibnamefont
  {Tesanovic}},\ }\href@noop {} {\bibfield  {journal} {\bibinfo  {journal}
  {Physical Review B}\ }\textbf {\bibinfo {volume} {87}},\ \bibinfo {pages}
  {104514} (\bibinfo {year} {2013}{\natexlab{b}})}\BibitemShut {NoStop}%
\bibitem [{ARG()}]{ARGO}%
  \BibitemOpen
  \href@noop {} {}\bibinfo {note} {Calculations were done on ARGO, a research
  computing cluster provided by the Office of Research Computing at George
  Mason University, VA. (URL: http://orc.gmu.edu)}\BibitemShut {NoStop}%
\bibitem [{\citenamefont {Fisher}\ and\ \citenamefont
  {Lee}(1989)}]{Fisher1989}%
  \BibitemOpen
  \bibfield  {author} {\bibinfo {author} {\bibfnamefont {M.~P.~A.}\
  \bibnamefont {Fisher}}\ and\ \bibinfo {author} {\bibfnamefont {D.~H.}\
  \bibnamefont {Lee}},\ }\href@noop {} {\bibfield  {journal} {\bibinfo
  {journal} {Physical Review B}\ }\textbf {\bibinfo {volume} {39}},\ \bibinfo
  {pages} {2756} (\bibinfo {year} {1989})}\BibitemShut {NoStop}%
\bibitem [{\citenamefont {Fisher}\ \emph {et~al.}(1989)\citenamefont {Fisher},
  \citenamefont {Weichman}, \citenamefont {Grinstein},\ and\ \citenamefont
  {Fisher}}]{Fisher1989a}%
  \BibitemOpen
  \bibfield  {author} {\bibinfo {author} {\bibfnamefont {M.~P.~A.}\
  \bibnamefont {Fisher}}, \bibinfo {author} {\bibfnamefont {P.~B.}\
  \bibnamefont {Weichman}}, \bibinfo {author} {\bibfnamefont {G.}~\bibnamefont
  {Grinstein}}, \ and\ \bibinfo {author} {\bibfnamefont {D.~S.}\ \bibnamefont
  {Fisher}},\ }\href@noop {} {\bibfield  {journal} {\bibinfo  {journal}
  {Physical Review B}\ }\textbf {\bibinfo {volume} {40}},\ \bibinfo {pages}
  {546} (\bibinfo {year} {1989})}\BibitemShut {NoStop}%
\end{thebibliography}

%

\end{document}